# Quantitative validation of SMBI deposition profiles on EAST using experimentally constrained simulations

Yangdi Yu[1,2], Tianyang Xia[*1], Jie Zhang[*2], Jia Huang[1], Huimin Qi[1,2], Dazheng Li[3], Yanlong Li[1], Xiaoxue He[1], Xiaowei Mao[2], Hang Yu[1,2], Shuang Zheng[1,2], Haoyu Nan[1,2], Junyi Yang[2], Kaining Lin[2] and the EAST team[†]

[1] Hefei Institutes of Physical Science, Chinese Academy of Sciences, Hefei 230031, P. R. China

[2] University of Science and Technology of China, Hefei 230026, P. R. China

[3] Dalian University of Technology, Dalian 116024, P. R. China

[*] Corresponding authors: (Tianyang Xia) xiaty@ipp.ac.cn, (Jie Zhang) jiez111@ustc.edu.cn

[†] See Ida et al. 2018 (https://doi.org/10.1088/1361-6587/aa9b03) for the EAST Team.

## Abstract

Quantitative comparison between SMBI fueling simulations and experiments has long been hampered by the poor constraint on the effective source. In this work, the particle increment per pulse is extracted experimentally and used to calibrate the simulation source. For EAST shot 71438, the density evolution measured by the microwave reflectometry is analyzed for nine groups of SMBI pulses. The density increase induced by SMBI is isolated from the background evolution by comparing the rising and falling phases around each pulse group. Its radial profile peaks in the mid-pedestal region for every pulse group. Volume integration of this increment provides the experimental reference used to calibrate the simulation source. The simulations are performed using multi-fluid model in BOUT++ framework in which two transport operators are recast into conservative flux form, with the residual of particle balance reduced to about 2% of the injected amount, and the perpendicular transport coefficients are constructed from the discrete radial flux balance, with the unphysical changes of the initial density and temperature profiles in source-free evolution reduced by an order of magnitude. For three cases sharing the same nominal molecular flux, the simulated deposition peak lies in the mid-pedestal region and agrees with the experimental one in the relative pedestal coordinate. The simulated conversion of the injected particles at the end of the injection is slightly lower than value inferred from the experiment, while the fraction retained at the end of the pulse is higher. The perturbation thus decays more slowly in the simulation than in the experiment, which we attribute to fast loss channels not represented in the model, such as the parallel losses to the scrape-off layer. The edge electron temperature stays on the high-temperature plateau of the dissociation and ionization rate coefficients throughout the pulse, and the inward shift of the deposition with edge cooling reported in earlier simulations does not occur under the present conditions.

## 1. Introduction

Plasma density control through fueling plays a major role in enabling long-term steady-state operation of future fusion reactors. Currently, the common fueling methods include gas puffing (GP), supersonic molecular beam injection (SMBI) [1], pellet injection (PI) [2], and compact toroid (CT) injection [3]. As a simple yet efficient fueling technique, SMBI has been successfully implemented on a wide range of devices, including EAST, Tore Supra, W7-AS, ASDEX-U, HL-2A, and others, yielding favorable results [1].

The fueling efficiency of SMBI is closely related to both the state of the background plasma and the parameters of the fueling system itself. It is generally recognized that a higher background temperature and density make fueling more challenging [4]. On the EAST tokamak, SMBI pulses of 10–15 ms inject molecules on the order of $10^{20}$ , and the fueling efficiency measured in ohmic discharges is 15%–30% [5].

Closing the particle balance in tokamaks requires resolving how the density, temperature and transport evolve on the time scale of a single SMBI pulse [6]. Such transient studies demand diagnostics with high temporal and spatial resolution. For the electron density, this capability is provided by microwave reflectometry (MRV), which can resolve the profile evolution on the time scale of an individual SMBI pulse.

While MRV can resolve the profile evolution on the time scale of individual SMBI pulse, the density profile alone does not directly reveal the underlying particle source and transport. During the density rise phase following SMBI, the profile shape is often constrained by profile stiffness, which makes it difficult to disentangle the fueling source from transport-driven redistribution. Moreover, neutrals from the injected beam can degrade the coherence of the reflectometry echo signals, further complicating the interpretation of the density measurement. Filling this gap requires forward modeling of the particle source and transport.

Therefore, simulations of the transport process following SMBI are an important complement to experiments. Commonly used codes include SOLPS, JOREK, and BOUT++ [7]. To gain further insight into the SMBI fueling process, several SMBI simulation codes based on the trans-neut prototype have been developed [8–11] and applied in a series of SMBI studies on HL-2A and EAST [12–15]. The simulated deposition location has been compared with the measured density shoulder on EAST [16], and the SMBI-induced changes of edge turbulence and of the confinement transition have been characterized experimentally on HL-2A [17].

A major obstacle to a quantitative comparison between simulation and experiment lies in the intensity of the source, which is difficult to determine from the available measurements. In many modeling studies, as a natural first approximation, all the molecules ejected from the nozzle are assumed to enter the plasma directly. In reality, a significant fraction is absorbed by the first wall, removed by pumping, or ionized in the scrape-off layer (SOL) and diverted, and never reaches the confined plasma. The effective source intensity, which is the number of molecules that actually penetrate the plasma, is therefore highly uncertain, and the codes are sensitive to it. Too many molecules can cool the edge and increase the penetration depth, while under certain conditions the self-shielding effect may instead reduce the penetration depth [18]. For codes such as trans-neut, an additional difficulty is that the required input parameters, namely the density, velocity, and poloidal width of the molecular beam, are rarely measured directly. What the EAST experiment typically provides is the Laval nozzle geometry and the backing pressure, from which the beam parameters must be inferred. With the effective source intensity poorly constrain and the beam parameters themselves uncertain, most simulations remain qualitative or semi-quantitative, and a direct quantitative comparison with experiment is correspondingly difficult.

The remainder of this paper is organized as follows. Section 2 introduces the experimentally inferred increment per pulse as the link between simulation and experiment, and this requires the code to conserve particle number. Section 3 extracts the particle increment per SMBI pulse from the reflectometry data. Section 4 describes the simulation setup, including the model, the injection parameters, and the conservative reformulation of the transport operators. Section 5 presents the simulation results and validates the simulated deposition against the experiment. Section 6 discusses the limitations of the model, and Section 7 summarizes the main conclusions.

## 2.Approach to quantitative validation

To bridge the gap quantitative between simulation and experiment, we take the experimentally measured change in the total particle number as the primary link between simulation and experiment. Making this link quantitative requires two ingredients. The first concerns the experiment. The part of the density evolution that is directly correlated with the SMBI pulses must be separated from the background evolution. We achieve this by statistically combining the many similar pulses within a single discharge. After the slowly varying background trend is removed, the rising and falling segments around each pulse group are fitted separately, and the difference between the two slopes isolates the decay-corrected density increase rate induced by SMBI, with the background contribution canceled out. The radial profile of this rate exhibits a distinct peak in the mid-pedestal region for every pulse group, which we interpret as the deposition of the injected particles. Volume integration of this rate in flux coordinates then gives the particle increment per pulse, which provides the experimental reference for the simulations. The second ingredient concerns the simulation. For the particle increment to serve as a quantitative link, the particle number in the simulation must itself be trustworthy. This imposes two requirements on the code.

The first requirement is that the transport operators conserve particle number on the discrete grid. The trans-neut code employs a magnetic flux coordinate system based on the BOUT++ framework [19] together with a finite-difference method formulated in that coordinate system [20]. The parallel advection is discretized with a third-order weighted essentially non-oscillatory (WENO) scheme, which, like other finite-difference schemes, is exactly conservative only for terms written in flux form [21]. Conservation must therefore be secured through the form of the operator. In the form in which these operators have been used in earlier applications, two of them account for almost all of the error. The advection term of the ion equation and the diffusion term of the atom equation are not written in the flux form required by the flux coordinate system, so that the geometric factors of the coordinate system are not accounted for and the particle number does not balance when it is summed over the grid with the correct volume weights. In this work both operators are recast into the corresponding conservative flux form. With them, the numerical closure of the particle budget is at the level of about two percent of the injected amount, and no a posteriori correction is required.

The second requirement is that the prescribed transport coefficients leave the initial profiles unchanged in the absence of a source. In the interpretive approach commonly adopted in BOUT++ transport modeling, the perpendicular transport coefficients are inverted from the measured profiles through the steady-state transport equations and then held fixed throughout the simulation [20,22,23]. The coefficients were previously inverted on the continuous steady-state equations, which is not consistent with the discretized operator actually used, so that even the source-free profiles were not strictly stationary. In this work, the coefficients are constructed directly from the discrete radial fluxes. The flux is required to be constant on each flux surface, and the local metric factor $Jg^{11}$ is retained cell by cell. The construction is described in Appendix 1. With it, the source-free evolution of the initial profiles is reduced by about an order of magnitude relative to the conventional inversion, and the constancy of the discrete flux provides a self-check of the coefficients that are actually used in the simulation. This construction determines only the radial and poloidal shape of the coefficients, Their absolute level remains a calibrated quantity, normalized at the inner boundary of the simulation domain. They are interpreted as an effective representation of the processes that sustain the H-mode background but are not modeled explicitly, an identification that relies on the pulse being short compared with the time scale on which that background evolves. Both elements are independent of the SMBI problem itself. They apply to any flux-coordinate transport code in which the transport coefficients are prescribed from experimental profiles.

We then perform simulations of a single pulse with the source calibrated by the experimentally inferred increment, for three combinations of molecular density and velocity.

## 3. Experimental data for inputs

In EAST, the time scale of a single SMBI pulse is typically on the order of milliseconds, while the time resolution of the MRV density diagnostic is on the order of 0.1 ms, so that the density ramp-up within a single pulse is resolvable in principle. In practice, however, random perturbations in the measurements and system response delays make this ramp-up difficult to be observed directly. By exploiting the large number of SMBI pulses with similar temporal structures delivered within one discharge, we employ statistical methods to separate the SMBI-induced evolution from the random perturbations, yielding a relatively accurate evolution of the density profile. This section describes the processing chain that yields the particle increment per SMBI pulse, which is used in Section 4 to calibrate the simulation source.

Shot 71438 is analyzed over the density ramp-up phase from t = 1.98 s to 2.50 s. The SMBI trigger signal, recorded with the frequency at 10 kHz, consists of a series of rectangular sub-pulses, each with an on-phase of about 1 ms followed by a pause of about 4 ms, appearing in nine groups (denoted SPG1–SPG9) comprising 33 sub-pulses in total, as listed in Table 1. The first and the last groups are statistically indistinguishable from the remaining groups and are therefore retained in the analysis.

| Group | Start time (s) | End time (s) | Number of sub-pulses |
|---|---|---|---|
| SPG1 | 1.9833 | 1.9993 | 4 |
| SPG2 | 2.0422 | 2.0582 | 4 |
| SPG3 | 2.1091 | 2.1252 | 4 |
| SPG4 | 2.1781 | 2.1942 | 4 |
| SPG5 | 2.2419 | 2.2531 | 3 |
| SPG6 | 2.2864 | 2.2979 | 3 |
| SPG7 | 2.3192 | 2.3255 | 2 |
| SPG8 | 2.3625 | 2.3847 | 5 |
| SPG9 | 2.4551 | 2.4712 | 4 |

**Table 1.** Start and end times and the numbers of sub-pulses of the nine SMBI pulse groups (SPG1–SPG9) in shot 71438. The groups are identified from the 10 kHz trigger signal.

Fig. 1 shows the density evolution at three representative radial positions, near the magnetic axis, at the pedestal top, and in the mid-pedestal, together with the SMBI pulse timing. Over the radial range measured by MRV the density rises during the pulse groups and falls in their absence, and the correlation with the SMBI pulse timing is clearest near the pedestal. The correlation is somewhat weaker for $R > 2.3$ m and $R < 2.1$ m, which may be attributed to the fact that the injected beam penetrates directly through the outermost low-temperature, low-density region and cannot effectively reach the magnetic axis.

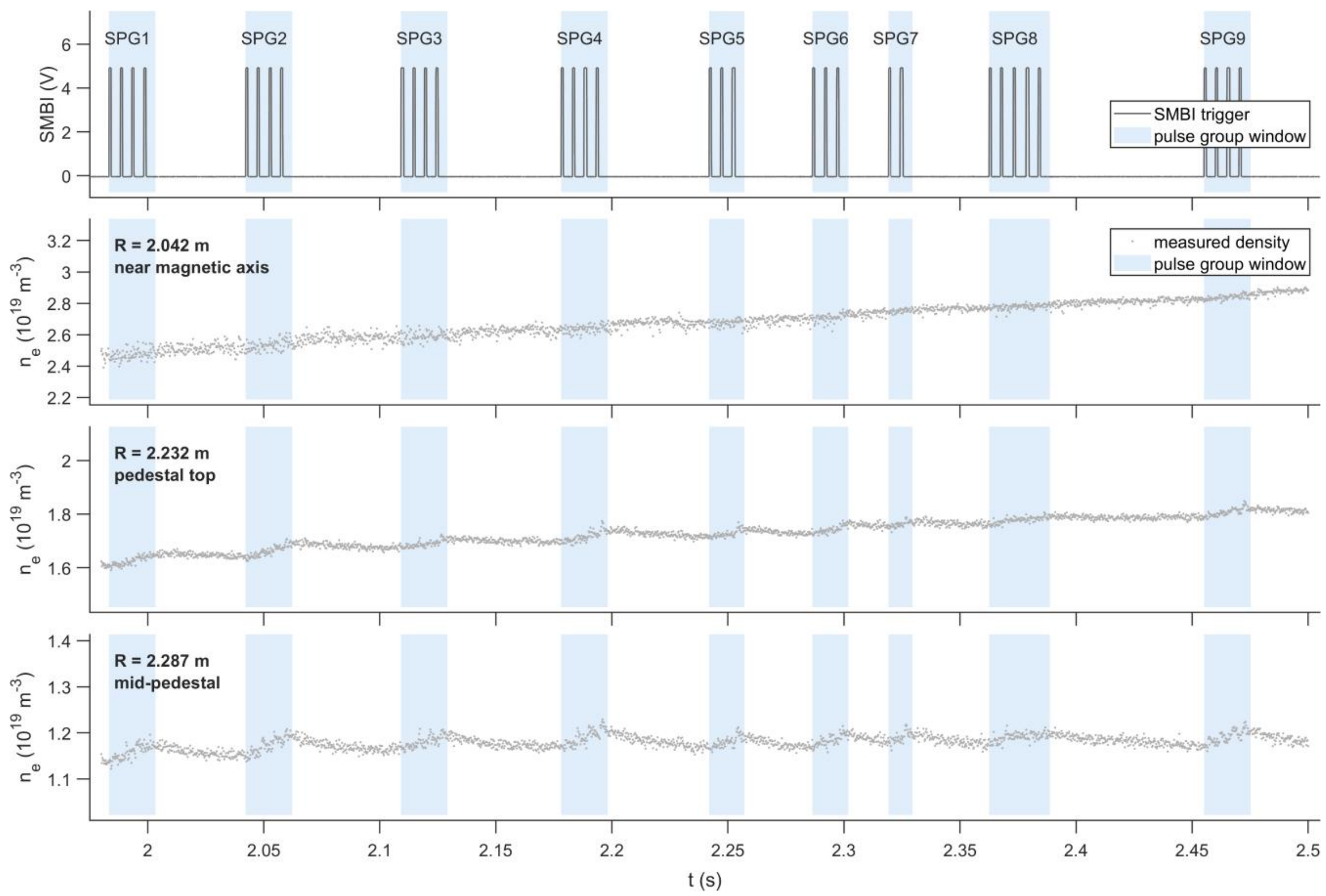

**Fig. 1.** Density evolution at three representative radial positions, near the magnetic axis (R = 2.042 m), at the pedestal top (R = 2.232 m), and in the mid-pedestal (R = 2.287 m), together with the SMBI trigger signal (top panel). The gray dots represent the measured density, and the shaded bands mark the nine pulse groups (SPG1–SPG9), spanning from the start of each group to 4 ms after its end. Near the pedestal, the density evolution shows a clear correlation with the SMBI pulse timing.

In addition, various factors can cause changes in the measured plasma density, such as turbulence-driven transport, heating-induced pressure variations, neutral beam injection, and even displacement of the plasma itself during the measurement. These contributions are not correlated with the SMBI pulse sequence, so the density change induced by SMBI can be isolated by comparing the evolution inside and outside the pulse intervals. Since the trigger signal marks the timing of the injection directly, whereas the density response is delayed and broadened by the plasma and by the diagnostic, these intervals are taken directly from the electrical trigger signal of SMBI rather than from the density response itself.

The processing chain consists of the following steps.

1. To obtain a regularly gridded dataset, the raw measurements are resampled onto 48 radial channels with a time resolution of 0.25 ms, yielding 2081 time points within the analysis window [1.98, 2.50] s and covering all nine pulse groups. The radial coordinate is interpolated and extrapolated onto a common radial axis spanning [1.92, 2.35] m with the reference axis taken as the time average of the measured channels. Since the boundary of the measured radial axis varies from time slice to time slice, the interpolated R coordinate may carry a small offset.

2. Based on the noise characteristics of the experimental data in the time-frequency domain, a median filter with a window size of 5 followed by a Gaussian convolution with a window size of 15 is applied to each radial channel to partially remove random perturbations. No smoothing is applied along the radial direction. The average density increase rate is then obtained by linear regression at each radial position. Fig. 2 demonstrates that this filtering improves the fitting confidence without affecting the fitting results.

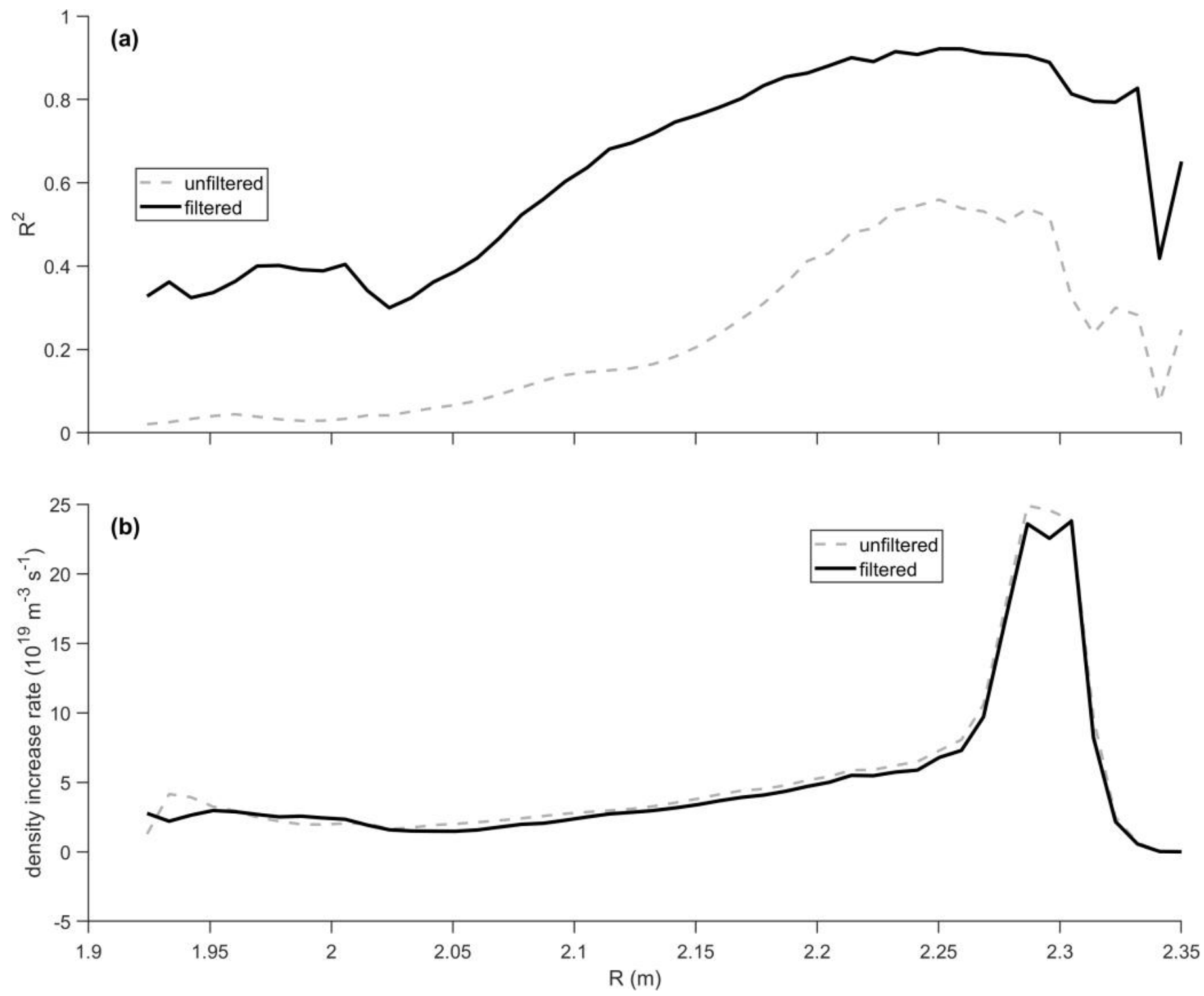

**Fig. 2.** Effect of the filtering on (a) the coefficient of determination $R^2$ of the linear regression and (b) the average density increase rate, both averaged over the nine pulse groups. Dashed gray curves: unfiltered data; solid black curves: filtered data (median filter of window 5 followed by Gaussian convolution of window 15). The filtering improves the fitting confidence without affecting the fitting results.

3. The trend of the average density increase is subtracted from the filtered data to obtain piecewise line segments with the average trend removed, which helps to exclude the effects of non-SMBI factors.

4. The resulting piecewise line segments are fitted segment by segment. For each pulse group, a rising window spanning the group plus 4 ms and a falling window spanning the following pause are identified. Linear fits in the two windows give the rising and falling slopes at each radial position. The rising slope contains both the SMBI source and the background loss, the falling slope reflects the background loss alone. So their difference $N_{\mathrm{fix}} = N_{rise} - N_{fall}$ gives the corrected density increase rate induced by SMBI, with the global trend canceled out. The fitting results at three representative positions are shown in Fig. 3. Additionally Fig. 4(a) shows the corrected density increase rate at different radial positions for the nine pulse groups, and Fig. 4(b) the corresponding coefficient of determination $R^2$.

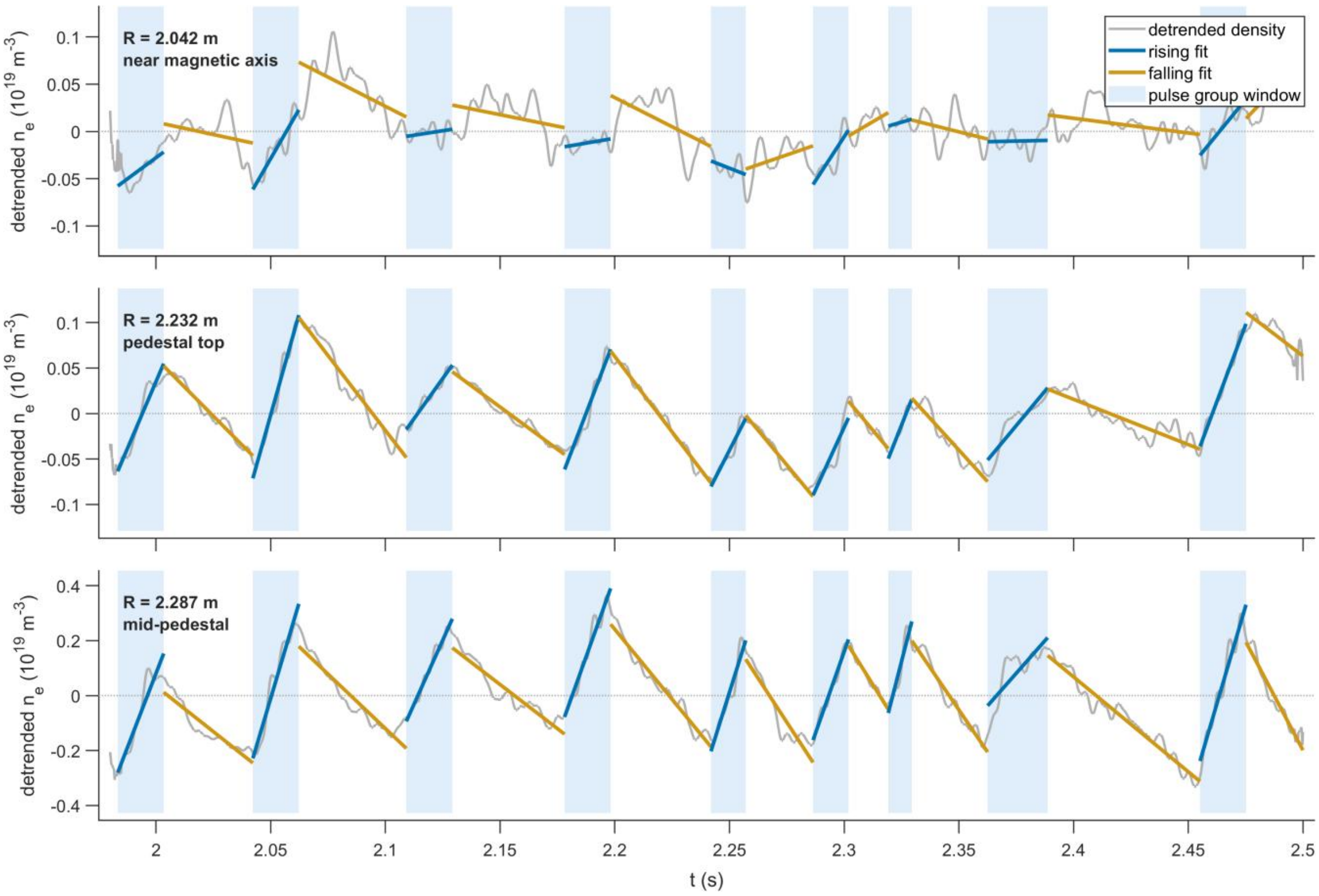

**Fig. 3.** Results of the piecewise fitting after detrending at the three representative radial positions of Fig. 1. The gray line represents the detrended density, the blue solid lines the linear fits of the rising windows, and the yellow solid lines those of the falling windows; the shaded bands mark the rising windows. The piecewise fitting separates the SMBI-induced density rise from the background decay.

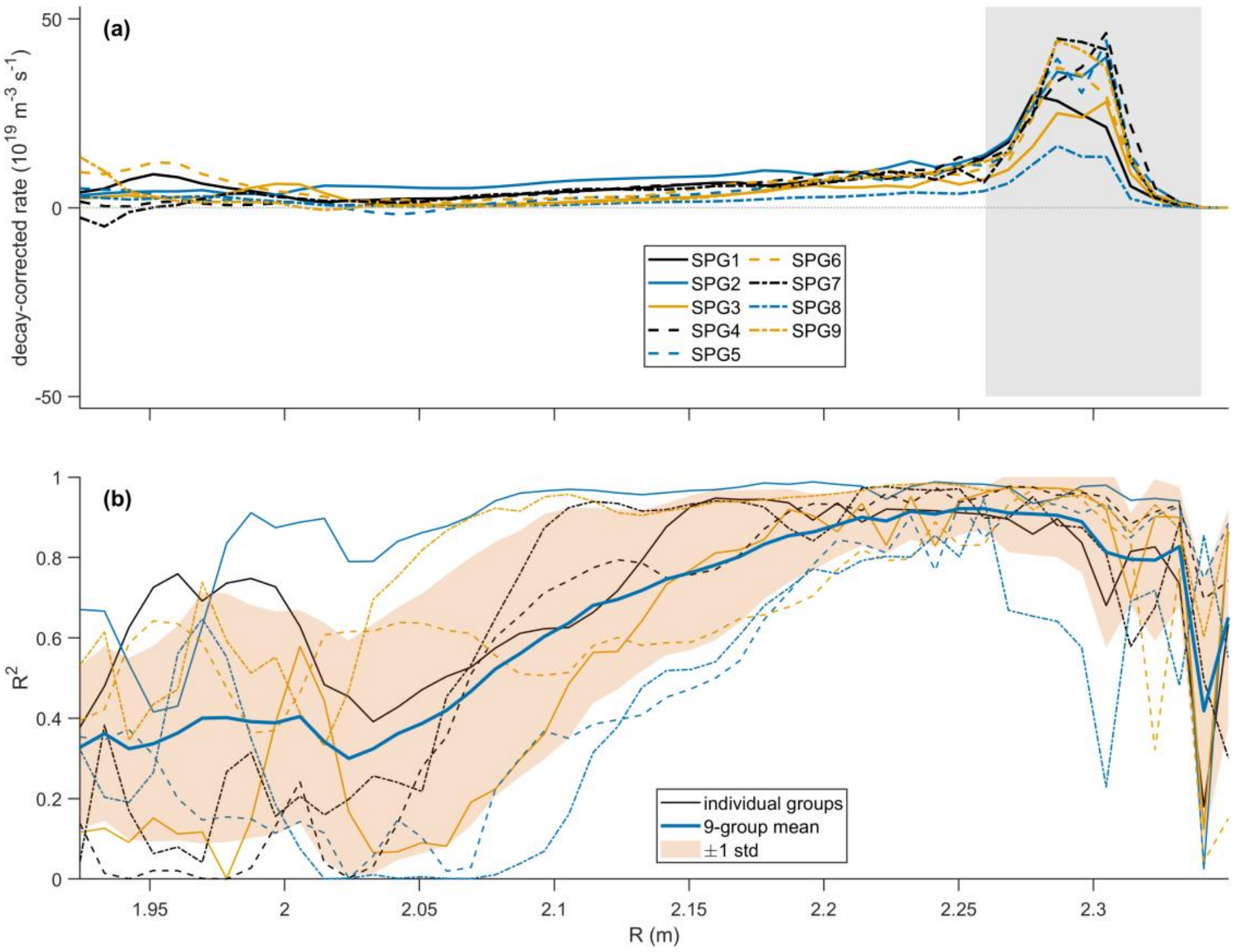

**Fig. 4.** (a) Radial distribution of the decay-corrected density increase rate for the nine pulse groups (solid lines: SPG1–SPG3; dashed lines: SPG4–SPG6; dash-dotted lines: SPG7–SPG9; black, blue, and orange within each triplet). The fastest density increase occurs in the mid-pedestal region (shaded band, R = 2.26–2.34 m) for all pulse groups. (b) Radial distribution of the coefficient of determination $R^2$ of the rising fits: the thin lines show the individual groups, the thick blue curve the average over the nine groups, and the red shaded band the ±1 standard deviation. The fitting quality is low near the magnetic axis, indicating a weak direct SMBI influence there.

5. The corrected density increase rate is multiplied by the 5 ms pulse period (1 ms on-phase followed by a 4 ms pause) to obtain the density change per pulse, and the particle increment per pulse is then obtained by volume integration. The volume element in the flux coordinates $(x, y, \zeta)$ is $\mathrm{d}V = (h_\theta/B_p)\,\mathrm{d}x\,\mathrm{d}y\,\mathrm{d}\zeta$, where $x$ and $y$ are the radial and poloidal coordinates, $\zeta$ the toroidal angle, $h_\theta$ the poloidal arc length metric, and $B_p$ the poloidal magnetic field. The integration over one full toroidal turn gives the form used below, $\mathrm{d}V = 2\pi(h_\theta/B_p)\,\mathrm{d}x\,\mathrm{d}y$. Since one hydrogen molecule yields two electrons, the molecular increment is half of the electron increment. Two integration domains are considered: the core inside the last closed flux surface ($\psi$ < 1) and the whole measurable volume. The results are shown in Fig. 5. The particle increment over the whole volume is $(0.424 \pm 0.037) \times 10^{19}$ molecules per pulse, within the core it is $(0.240 \pm 0.021) \times 10^{19}$ molecules per pulse. The uncertainties are standard errors of the mean over the nine pulse groups, for which the group-to-group scatter is about $\pm$26% in both domains. About 7% of the positive source located beyond the outermost grid radius of the integration mesh (R ≈ 2.31 m) is not included.

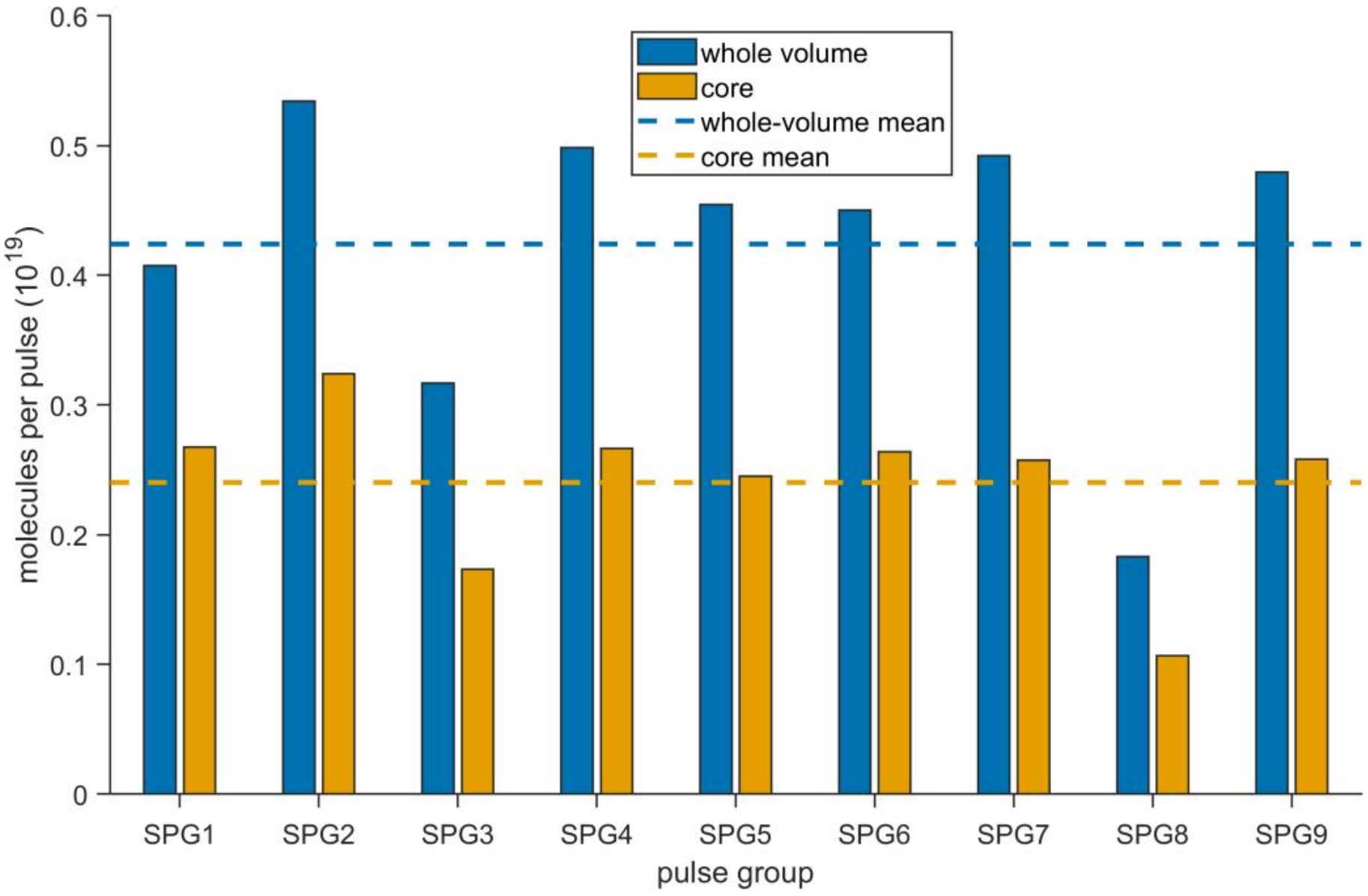

**Fig. 5.** Particle increment per SMBI pulse (average over the 5 ms pulse period) for each of the nine pulse groups, computed by volume integration with the flux-coordinate volume elements $\mathrm{d}V = 2\pi(h_\theta/B_p)\,\mathrm{d}x\,\mathrm{d}y$. Blue bars: whole measurable volume; orange bars: core (inside the last closed flux surface within the main poloidal band). The dashed lines indicate the averages over the nine groups: $0.424 \times 10^{19}$ (whole volume) and $0.240 \times 10^{19}$ (core) molecules per pulse. One molecule corresponds to two electrons.

While the average increase rate of Fig. 2(b) reflects the overall density evolution, the decay-corrected rate of Fig. 4(a) isolates the part that is directly correlated with the SMBI pulse signal. In all nine pulse groups, the region exhibiting the fastest density increase is located in the mid-pedestal region. This is also where the fit quality is highest (Fig. 4(b)). Near the magnetic axis, the fit quality is low while the average increase rate is not lower than elsewhere (Fig. 2(b)), suggesting that the direct influence of SMBI near the magnetic axis is weak and that the density increase there is mainly caused by other effects. Since the R coordinate is obtained by interpolating the raw measurements, a small offset may exist, so the relative position between the density increase region and structures such as the pedestal should be interpreted with care.

The particle increments obtained above, $0.424 \times 10^{19}$ molecules per pulse over the whole volume and $0.240 \times 10^{19}$ within the core, provide the experimental reference for calibrating the molecular source of the simulations in Section 4, where the injection is prescribed by pinning $N_m$ and $V_{mx}$. In the experiment, 94.2% of the increment over the whole volume and 94.3% of that inside the core are still retained at the end of the injection, while the corresponding fractions 4 ms after the injection ceases are 71.2% and 71.4%. The retention is discussed together with the simulation results in Section 5.2.

# 4. Simulation settings

## 4.1 Simulation model and geometry

The trans-neut code, developed within the BOUT++ framework [19] and widely used for SMBI simulations [8–11], tracks four elementary processes following SMBI, namely the dissociation of injected molecules into atoms, ionization of atoms into ions, recombination of ions into atoms, and charge exchange between ions and atoms. The molecular beam is treated as a fluid, and its transport along the injection direction is solved together with the parallel and perpendicular transport of ions and atoms. Except where stated otherwise, the implementation details and the values of the model constants not specified in this section are those of Wang et al. [8] and Li et al. [20].

Simulations are performed on a 2.5D flux-coordinate grid with 128 radial and 512 poloidal points (132 radial points including the guard cells), covering the normalized poloidal flux range $\psi \in [0.55,0.99]$ (Fig. 6), in the same normalization as in Section 3, with $\psi = 1$ at the last closed flux surface. The magnetic equilibrium is reconstructed from EAST shot 71438 with EFIT, and the initial density and temperature profiles are mapped onto the same flux coordinates. The equilibrium and the initial profiles are taken from a later, flat-top stage of the discharge, at $t$ = 5.059 s, whereas the experimental analysis covers the density ramp-up phase. The two stages differ only slightly in their plasma parameters, and they remain representative of the discharge.

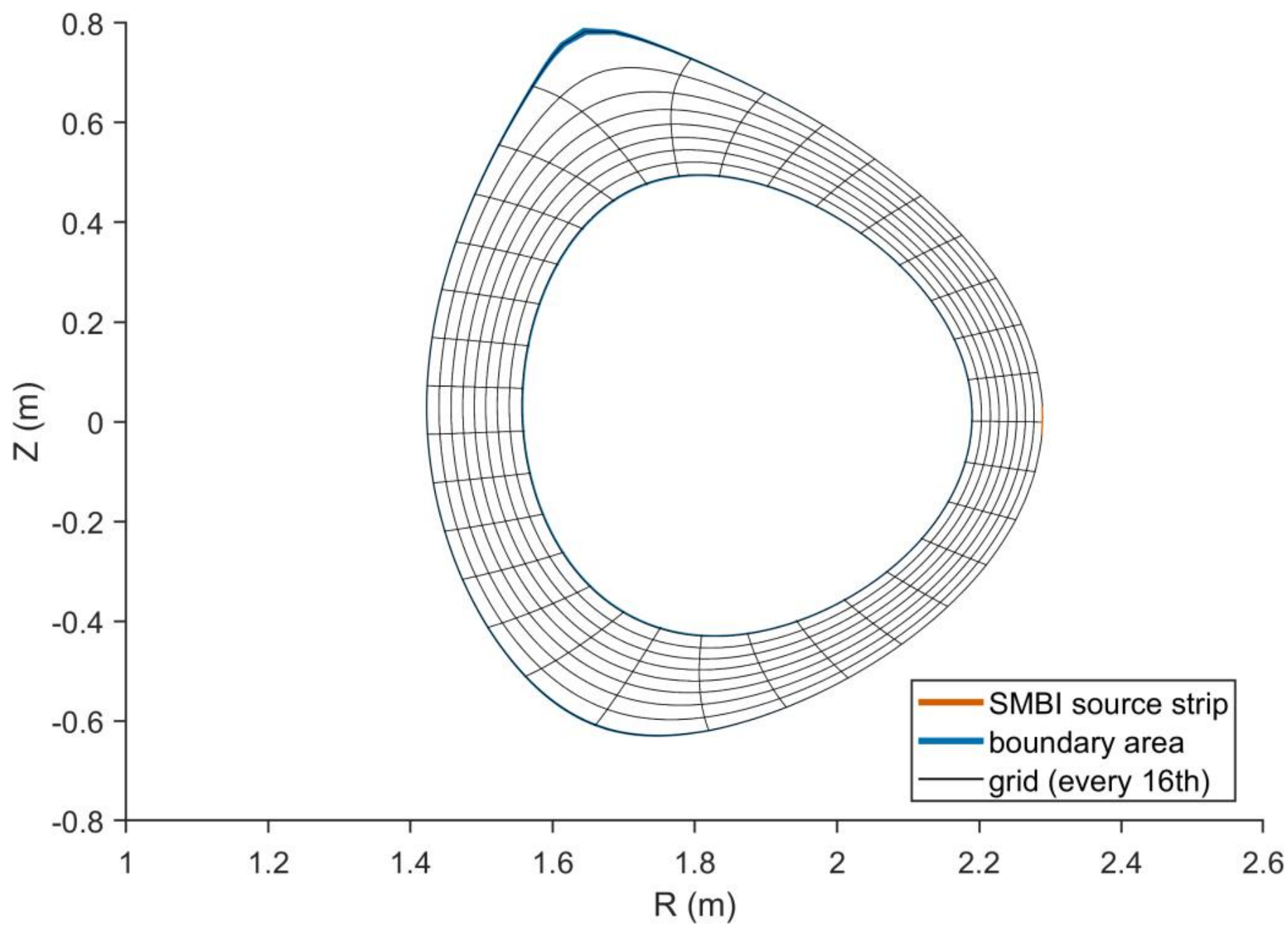

**Fig. 6.** The 2.5D flux-coordinate simulation grid (128 radial × 512 poloidal points, 132 radial points counting the guard cells, $\psi \in [0.55,0.99]$) in the poloidal plane. The blue bands mark the radial boundary regions, and the red strip at the OMP is the SMBI source region, where $N_m$ and $V_{mx}$ are pinned. Only every 16th grid line is drawn.

The background equilibrium is representative of an H-mode discharge. The perpendicular transport coefficients are constructed to maintain the initial profiles in the absence of the source. The duration of the pulse is assumed to be short compared with the time scale on which the background evolves. The ion inventory injected per pulse corresponds to about 5% of the plasma ion content within the simulated domain, and the resulting peak increase of the ion density reaches about 19% of the pedestal-top density. The coefficients are therefore treated as time-independent and applied unchanged during the injection. The stationary effect may exist is contained in the coefficients, what is not represented is their response to the fueling perturbation. Events that can relax the pedestal on a millisecond time scale, such as ELM crashes (whose effects on the pedestal profiles recover within 3–8 ms at ASDEX Upgrade [24]), may occur within the analyzed window and are not described by the model. Their possible influence is discussed in Section 6.

The perpendicular transport coefficients are not part of the original trans-neut model. In earlier applications they were inverted from the initial profiles on the continuous steady-state equations [20]. Here they are constructed in the discrete flux form of the numerical scheme. On each flux surface the radial flux is required to be constant, and the local metric factor ($Jg^{11}$) is retained cell by cell, so that the discrete flux balance is closed by the coefficients actually used in the simulation. Their absolute value is a calibrated quantity, normalized at the inner boundary of the domain, whose construction is described in Appendix 1.

The particle transport of the three species is governed by the following equations. The poloidal advection operator in Eq. (1) and the perpendicular diffusion operator in Eq. (2) are written in conservative flux forms, which differ from the formulation used in the earlier applications [8–11]. The subscripts ∥ and ⊥ denote the components parallel and perpendicular to the magnetic field, with $\nabla_{\parallel} f = (\boldsymbol{b}\cdot\nabla)f$ and $\nabla_{\perp} f = \nabla f - (\boldsymbol{b}\cdot\nabla)f$, where $\boldsymbol{b} = \boldsymbol{B}/B$.

$$\frac{\partial N_i}{\partial t} = \nabla_{\perp}\cdot(D_{\perp}\nabla_{\perp}N_i) - \frac{1}{J}\frac{\partial}{\partial y}\left(\frac{J V_{\parallel i} N_i}{\sqrt{g_{22}}}\right) + S_{Ip} - S_{\mathrm{recp}} \qquad (1)$$

$$\frac{\partial N_n}{\partial t} = \nabla_{\parallel}\cdot(D_{\parallel}\nabla_{\parallel}N_n) + \nabla_{\perp}\cdot(D_{\perp}\nabla_{\perp}N_n) - S_{Ip} + S_{\mathrm{recp}} + 2S_{\mathrm{dissp}} \qquad (2)$$

$$\frac{\partial N_m}{\partial t} + \nabla_x(V_m N_m) = -S_{\mathrm{dissp}} \qquad (3)$$

$$\frac{\partial V_m}{\partial t} + V_m\nabla_x V_m = -\frac{\nabla_x P_m}{N_m M_m} \qquad (4)$$

where $N_i$, $N_n$, and $N_m$ are the ion, atom, and molecular densities, respectively, $V_{\parallel i}$ is the ion velocity along the magnetic field, $V_m$ is the molecular velocity along the injection direction $x$, $y$ is the poloidal (parallel) coordinate, $J$ and $\sqrt{g_{22}}$ are the Jacobian and the poloidal arc-length metric of the flux coordinates, and $D_{\perp}$ and $D_{\parallel}$ are the perpendicular and parallel diffusion coefficients of the respective species: $D_{\perp}$ for the ions in Eq. (1), and $D_{\parallel}$ and $D_{\perp}$ for the atoms in Eq. (2). Quasi-neutrality ( $N_i = N_e$ ) is assumed, so that no separate electron density equation is required. The molecular pressure is $P_m = T_m N_m$ with the molecular temperature fixed at room temperature $T_m$ = 300 K, and $M_m$ is the molecular mass. The momentum equation of the molecular beam, Eq. (4), is shown here because the beam is treated as a fluid, and its velocity is quantitatively prescribed by the injection setup of Section 4.2. The source terms are defined as

$$S_{\mathrm{recp}} = N_i^2\langle\sigma_{\mathrm{rec}}V_{th,e}\rangle, \quad S_{Ip} = N_i N_n\langle\sigma_I V_{th,e}\rangle, \quad S_{\mathrm{cxp}} = N_i N_n\langle\sigma_{CX}V_{th,i}\rangle, \quad S_{\mathrm{dissp}} = N_i N_m\langle\sigma_{\mathrm{diss}}V_{th,e}\rangle \qquad (5)$$

where $\sigma_{\mathrm{rec}}$, $\sigma_I$, $\sigma_{CX}$, and $\sigma_{\mathrm{diss}}$ are the cross sections of recombination, ionization, charge exchange, and dissociation, respectively, $V_{th,e}$ and $V_{th,i}$ are the electron and ion thermal velocities, and $\langle\cdot\rangle$ denotes the average of the cross section times the relative velocity over the velocity distributions of the two reacting species. For the electron-impact processes the relative velocity is dominated by the electron thermal velocity, whereas for charge exchange it is that of the ions. The charge-exchange source $S_{\text{cxp}}$ conserves the total particle number and enters the temperature and momentum equations rather than the continuity equations. It is defined here for completeness. The equations of the ion velocity and of the electron and ion temperatures, together with the reaction-rate fits, are given in Appendix 1.

To close the system, boundary conditions are chosen according to three principles, namely consistency with the physical model, maintenance of the profiles in the absence of SMBI, and Neumann conditions by default. Dirichlet conditions take the value of the corresponding profile at the boundary, and Neumann conditions its derivative there. The complete set is listed in Appendix 3. Combined with the adopted transport coefficients, this setup keeps the density and temperature profiles quasi-steady in the absence of SMBI. Over the 1 ms source-free stage the maximum and volume-weighted changes of the three profiles are listed in Table 2. The changes presented below are therefore caused by the injection and the transport it induces.

| profiles | peak change (%) | volume-weighted mean (%) |
|---|---|---|
| Ni / Ti / Te (this work) | 0.06 / 0.19 / 0.43 | 0.001 / 0.007 / 0.10 |
| Ni / Ti / Te (continuous inversion) | 0.93 / 2.22 / 3.60 | 0.05 / 0.45 / 1.03 |

**Table 2.** Relative changes of the ion density, ion temperature and electron temperature over the 1 ms source-free stage in the simulations, for the transport coefficients constructed in this work and for coefficients inverted on the continuous equations [20]. Peak values are the largest over the 128 radial grid points, and volume-weighted values are averaged with the volume of the flux shell as weight. The largest changes occur near the inner and outer boundaries of the domain.

## 4.2 SMBI source and injection parameters

The aim of the injection setup is to prescribe a fixed molecular flux into the domain. Following the implementation of the code in the earlier applications [8–11], the injection is described by pinning $N_m$ and $V_{mx}$ in the source strip at the outer mid-plane (OMP) (Fig. 6), so that in the actual runs the molecular flux entering the domain beyond the source strip deviates slightly from the nominal value. The three cases, listed in Table 3, share the same nominal molecular flux $N_m V_m$ = 3 × $10^{21}$ m$^{-2}$ s$^{-1}$.

| Case | $N_m$ ($10^{19}$ m$^{-3}$) | $V_m$ (m s$^{-1}$) | $N_m V_m$ ($10^{21}$ m$^{-2}$ s$^{-1}$) |
|---|---|---|---|
| case 1 | 1.0 | 300 | 3.0 |
| case 2 | 0.5 | 600 | 3.0 |
| case 3 | 0.3 | 1000 | 3.0 |

**Table 3.** Injection parameters of the three simulation cases. The nominal molecular flux $N_m V_m$ is the same for all cases.

The molecular velocities $V_m$ listed in Table 3 are physical velocities. In the code, the value used as the source strip is $V_{mx}$, the contravariant velocity in the normalized flux coordinates. They are related by:

$$V_m = V_{mx} \frac{\bar{L}\, b_{\text{mag}}}{R\, B_{pxy}}$$

which for $\bar{L} \approx R$ reduces to $V_m \approx V_{mx}\, b_{\mathrm{mag}}/B_{pxy} \approx 14.36\, V_{mx}$ at the injection strip, where $b_{\mathrm{mag}}$ is the magnitude of the magnetic field, $B_{pxy}$ the poloidal magnetic field, $R$ the major radius, and $\bar{L}$ the reference length. Together with the strip area of 0.896 $m^2$ ,the calibrated injection realized in the runs is $0.282 \times 10^{19}$ molecules per pulse.

Each simulation runs for 6 ms in total and is divided into three stages, namely the pre-injection stage of 1 ms with the source switched off, which establishes the unperturbed profiles, the injection stage of 1 ms, and a post-injection stage of 4 ms. The last two stages together constitute one complete pulse of 5 ms, matching the pulse duration of the discharge analyzed in Section 3.

Although the source is thus prescribed in absolute terms, the source intensity is not intended to reproduce the absolute value of the experimental particle increment. In this methodology its value is fixed by the calibration against the experimental increment, so that it is not an independently determined quantity, which is not used as a validation target, and only dimensionless ratios are compared with experiment. As shown in Fig. 7, the theoretical injection, $0.564 \times 10^{19}$ ion equivalents per pulse, lies between the experimental per-pulse increments inside the core ($\sim 0.48 \times 10^{19}$ ions) and over the whole measurable volume ($\sim 0.85 \times 10^{19}$ ions). The cumulative molecular dissociation within the injection stage closes to 95–101% of the theoretical injection for all three cases.

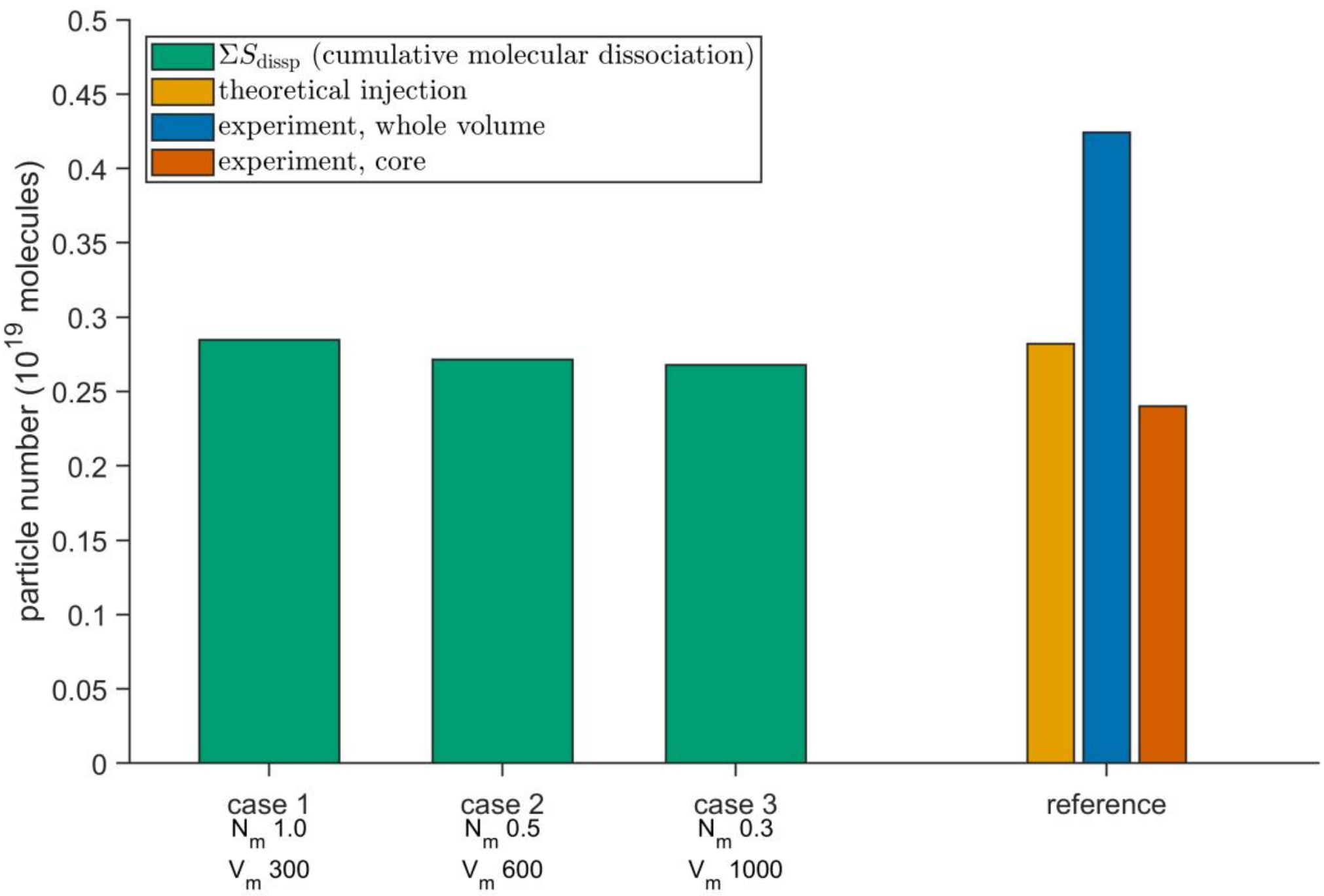

**Fig. 7.** Cumulative molecular dissociation within the injection stage (green bars) for the three cases, together with the theoretical injection (orange) and the experimental per-pulse particle increments over the whole volume (blue) and inside the core (red). The dissociation closes to 95–101% of the theoretical injection, and the theoretical injection falls between the experimental core and whole-volume increments.

## 4.3 Particle number conservation

Because the particle increment serves as the bridge between simulation and experiment, the simulated source is calibrated against the experimentally inferred increments, and the comparisons in Section 5.2 rely on particle counts, the accuracy of the particle accounting sets the level of the comparisons made below. We therefore examine the particle-number conservation of the transport operators in detail.

A term-by-term particle budget shows that the non-conservation arises predominantly from two operators. They are the poloidal advection term of the ion continuity equation and the perpendicular diffusion term of the atom equation, whereas the remaining processes contribute only marginally. Both defects stem from the operator form rather than from the discretization. In flux coordinates, the poloidal volume element varies along the poloidal direction ($\sqrt{g_{22}}$ is not constant), so that an advection operator written in the non-conservative form $V_{\parallel}\nabla_{\parallel}N$ does not preserve the poloidally summed quantity $\sum_y N\sqrt{g_{22}}$ and generates a significant spurious ion accumulation. We therefore employ the conservative flux form shown in Eq. (1), $(1/J)\,\partial_y(JV_{\parallel}N/\sqrt{g_{22}})$, which reduces to boundary terms when summed with the volume weight $J$. Similarly, a perpendicular diffusion operator with the coefficient outside the divergence, $D\nabla_{\perp}^2 N_n$, does not admit a discrete flux balance in the curved metric, since its cell-integrated divergence does not vanish identically, and atoms are spuriously created along the beam channel. We therefore employ the flux form $\nabla_{\perp}\cdot(D_{\perp}\nabla_{\perp}N_n)$ shown in Eq. (2). The discretized expressions of the two reformulated operators are listed in Appendix 1.

With these reformulations, the cell-by-cell closure residual of the particle budget is reduced to about 2% of the injected amount over the injection stage. It remains at this level over the complete pulse and vanishes in the source-free stage. Fig. 8 shows the particle budget of one case (case 2) over a complete pulse. During the injection window the ion increment follows the cumulative dissociation closely, and after the injection the ion decay is dominated by the diffusive loss, whose slope matches that of the decay. The poloidal advection contribution remains near zero throughout, and the cumulative recombination is minor. In the post-injection stage, the accumulated dissociation and ionization get saturated, and no spurious source appears, with the closure residual over the complete pulse remaining on the order of 2% of the injected amount. Conservation is therefore enforced at the operator level.

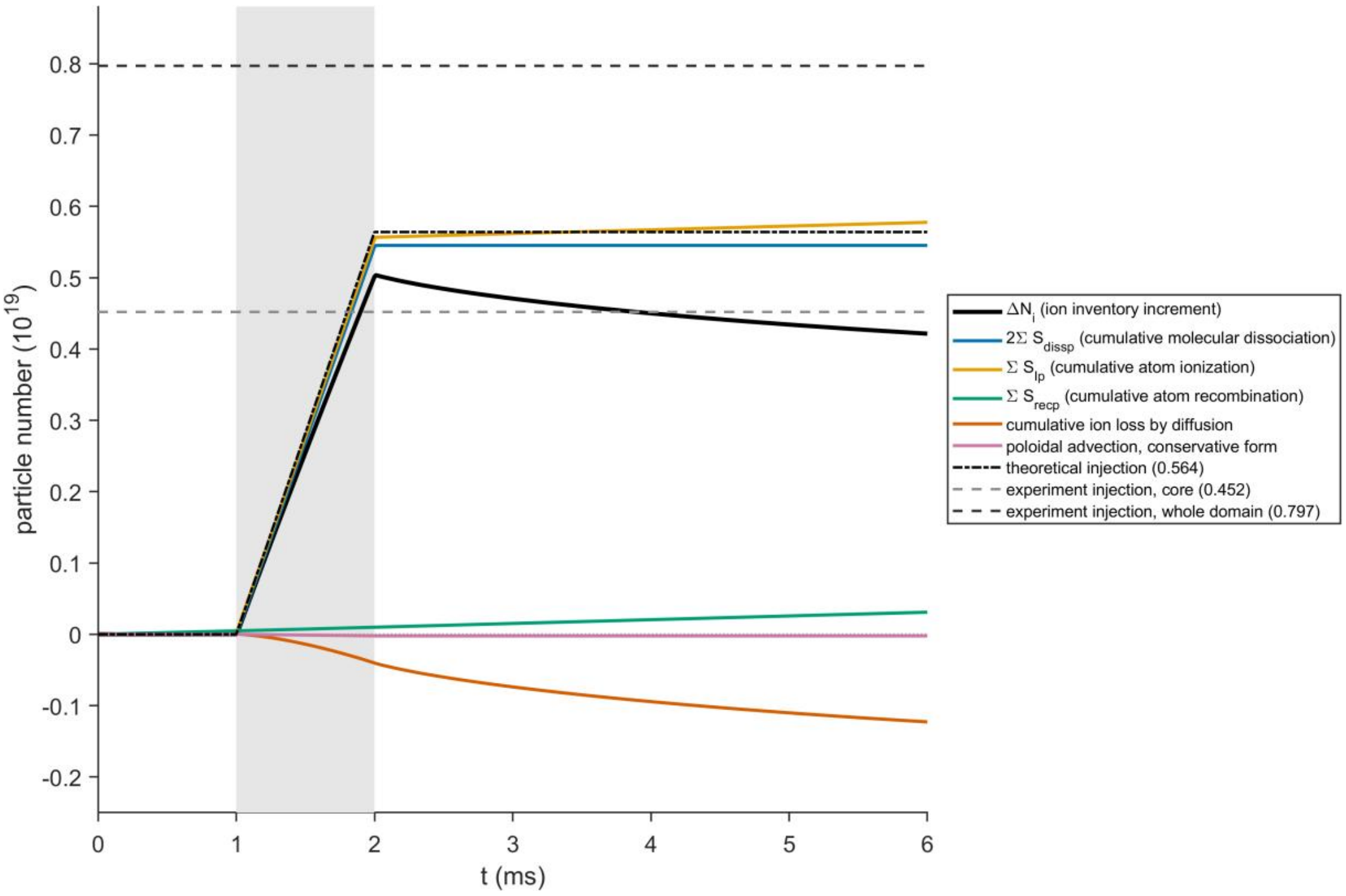

**Fig. 8.** Particle-number budget of case 2 over one complete pulse. The shaded band marks the injection window (1–2 ms). $\Delta N_i$ is the ion inventory increment. $2\Sigma S_{\mathrm{dissp}}$, $\Sigma S_{Ip}$, and $\Sigma S_{\mathrm{recp}}$ are the cumulative molecular dissociation (in ion equivalents), atom ionization, and recombination, respectively. The red and purple curves are the cumulative contributions of the perpendicular diffusion and of the poloidal advection to the ion balance. The dash-dotted line is the theoretical injection, and the dashed lines are the experimental per-pulse increments inside the core and over the whole domain. After the injection, $\Delta N_i$ decays with a slope matching the diffusive loss, and the advection contribution stays near zero, indicating conservative behavior.

We further assessed the influence of each modification on the particle accounting. The advection reformulation removes the dominant spurious contribution. After that reformulation, the remaining closure error is dominated by the diffusion operator of the atom equation, whose reformulation brings the residual down to the level quoted above. The other reformulations, such as heat conduction, viscosity, and the neutral poloidal diffusion, have only a minor impact on the particle accounting. All particle-number comparisons in the following sections are therefore accurate to about 2% of the injected amount.

# 5. Simulation results

## 5.1 SMBI process

The simulation setup keeps the density and temperature profiles approximately constant in the absence of SMBI, since the boundary conditions and the transport coefficients are constructed from the initial profiles. All the changes observed below can therefore be attributed to the injection and to the transport.

We first follow the injected particles through the pulse. At the OMP of the injection location, the ion density rises slowly during the injection (t = 1–2 ms) and decays after it stops (t = 2–6 ms), while the region of enhanced density spreads from the pedestal top towards both sides (Fig. 9(a)). The molecules did not arrive there directly. They travel along the injection direction and dissociate into atoms. The atoms diffuse both radially and poloidally and are ionized within a short distance, with a small fraction returns to the atomic state through recombination. The molecular and atom densities therefore reach a quasi-steady state shortly after the injection begins, whereas the ion density rises gradually with the injection (Fig. 9(b), (c)). The ionization source has the functional form $S_{Ip} \propto N_n N_i$. As the injection raises the atom density and the ion density together, the ionization rate first rises and then falls along the radial direction from the edge inward, so that the newly created ions are deposited in the pedestal. There the deposition is further shaped by the perpendicular diffusion coefficient $D_\perp$, which is lower where the pedestal gradient is steep. The ions therefore accumulate in the mid-pedestal region. Turbulent transport induced by the injection is not contained in the model, and its possible influence is discussed in Section 6. The molecular velocity V_m stays approximately unchanged throughout (Fig. 9(f)). Because the pressure term of the molecular momentum equation contains only $N_m$ and it is spatially smooth along the beam path and quasi-stationary in time. Inward of the deposition region $N_m$ falls rapidly to a very low level, where the fluid description of the molecules is no longer reliable. So that $V_m$ has no physical meaning there. This limitation applies to the molecular velocity only.

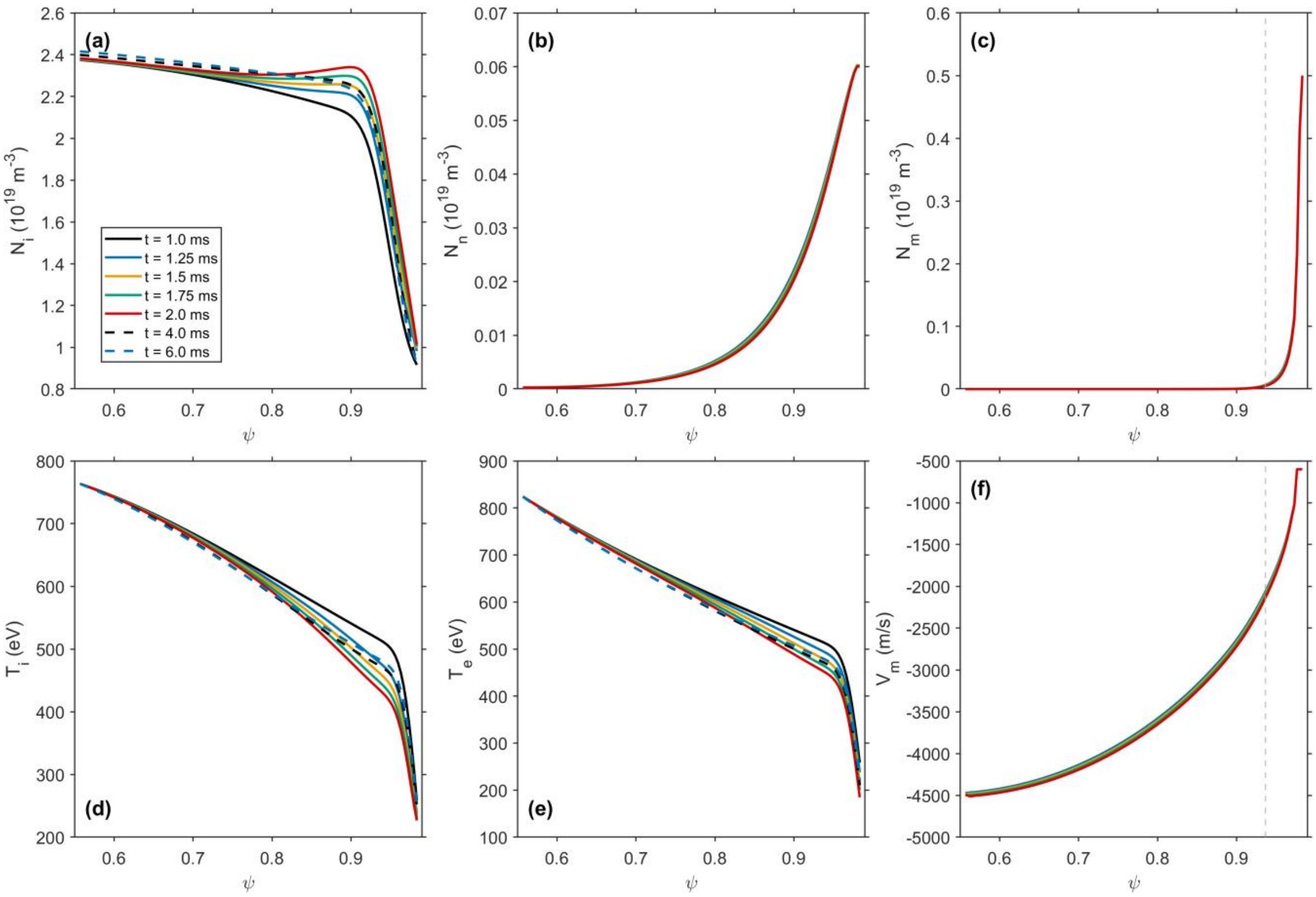

**Fig. 9.** Evolution of the radial profiles at the OMP of the injection location for a representative case (case 2). The horizontal axis is the normalized poloidal flux $\psi$. (a) ion density $N_i$, (b) atom density $N_n$, (c) molecular density $N_m$, (d) ion temperature $T_i$, (e) electron temperature $T_e$, (f) molecular velocity $V_m$, whose sign denotes the direction of motion, a negative value being directed inward. Solid lines are the injection stage ($t$ = 1.0/1.25/1.5/1.75/2.0 ms) and dashed lines are the post-injection stage ($t$ = 4.0/6.0 ms). $N_n$, $N_m$ and $V_m$ have a physical meaning during the injection only, and four instants of the injection stage are drawn for them. The vertical dashed line in (c) and (f) marks $\psi$ = 0.9367, where $N_m$ = 1% $N_{m0}$, and is a reading aid only; it does not represent a physical boundary.

The dissociation source $S_{\text{dissp}}$ peaks at the outer edge ($\psi$ = 0.9799) and is quasi-stationary during the injection. This shows that the molecules hardly penetrate into the high-parameter plasmas and the dissociation is concentrated at the edge. The deposition peak in the pedestal ($\psi$ = 0.9463) is therefore built by the inward transport of the atoms and ions, consistent with the path described above. Whether the neutrals can penetrate deeper is controlled by the edge temperature. Both the ion and the electron temperature fall during the injection and recover afterwards, and the temperature perturbation is confined to the pedestal, with little effect further inward (Fig. 9(d), (e)). Throughout the injection Te on the injection trajectory stays more than an order of magnitude above 50 eV. The dissociation and ionization coefficients vary rapidly under 50 eV (Fig. 12). Only when the edge temperature falls into this range and the rates are substantially reduced, the neutrals are able to penetrate deeper into the plasmas and the deposition ia able to move inward. That condition does not occur in these EAST simulations, so that the deposition peak remains in the pedestal region. Therefore, the inward motion of the density peak with decreasing temperature reported in Ref. [8,9] is not found here.

The two temperatures fall almost in step at the pedestal, but this is not a consequence of the electron-ion energy exchange. With the collision time of Eq. (9), the electron-ion energy equipartition time, defined by $\partial T_e/\partial t = -(T_e - T_i)/\tau_{\text{eq}}$, is $\tau_{\text{eq}} = M_i\tau_{ei}/(2m_e) = 1836\,\tau_{ei} \approx 25$ ms in the pedestal, which is far longer than the injection. So the exchange term is negligible. The real driving term is the ionization sink of identical structure in the two temperature equations, $-\nu_{\mathrm{ionz}}T$. The newly ionized ions dilute the ion temperature, and each ionization also removes energy from the electron channel. Both equations therefore give the same relative cooling rate, $\mathrm{d}(\ln T)/\mathrm{d}t = -\nu_{\mathrm{ionz}}$, and since $T_i \approx T_e$ at the pedestal the changes in the two temperatures are also similar. In addition, the ion temperature equation contains a term that has no counterpart in the electron equation, the compression term of the velocity field, $-\frac{2}{3}T_i\nabla\cdot\boldsymbol{V}$, whose magnitude is about 0.65 times of the dilution term (Table 4). the effect of its absence from the electron equation is discussed in Section 6. The absolute difference between the two temperatures (about 61 eV) is mostly the difference between the two core boundary condition (62.9 eV) rather than a dynamic separation produced by the exchange.

| $\psi$ | species | dilution ($\mathrm{eV\,ms^{-1}}$) | compression ($\mathrm{eV\,ms^{-1}}$) | parallel conduction ($\mathrm{eV\,ms^{-1}}$) | other ($\mathrm{eV\,ms^{-1}}$) | net $\mathrm{d}T/\mathrm{d}t$ ($\mathrm{eV\,ms^{-1}}$) |
|---|---|---|---|---|---|---|
| 0.933 | Ti | −5278.2 | −3453.2 | 8615.5 | 34.3 | −81.6 |
| 0.933 | Te | −5479.4 | – | 5531.1 | −114.8 | −63.2 |
| 0.946 | Ti | −6165.7 | −4037.4 | 10030.5 | 84.6 | −88.0 |
| 0.946 | Te | −6451.4 | – | 6512.5 | −127.7 | −66.6 |
| 0.973 | Ti | −6275.6 | −4077.5 | 10175.7 | 91.5 | −85.9 |
| 0.973 | Te | −6602.8 | – | 6843.1 | −318.8 | −78.5 |
| 0.983 | Ti | −4770.7 | −2706.5 | 7184.7 | 258.4 | −34.0 |
| 0.983 | Te | −4436.7 | – | 4680.1 | −322.2 | −78.7 |

**Table 4.** Mean rates at which the energy terms change the ion and electron temperatures over the injection stage ($t$ = 1 → 2 ms). The unit is $\mathrm{eV\,ms^{-1}}$ and negative values denote cooling. As the stage lasts 1 ms, each entry is numerically equal to the contribution integrated over the stage in eV. Positions are labeled by the normalized poloidal flux $\psi$ and are taken at the OMP. The dilution term $-\nu_{\mathrm{ionz}}T$ is common to the two temperature equations, whereas the compression term $-\frac{2}{3}T_i\nabla\cdot\boldsymbol{V}$ exists in the ion equation only. The column "other" collects the net contribution of the perpendicular diffusion and other terms for the ions, and mainly the ionization and dissociation energy costs for the electrons. The last column is the mean rate of change of the simulation temperature over the stage. Each term is given by an accumulator integrated in step with the temperature equation, and the terms sum to the simulation value to within a relative residual below $10^{-4}$ before rounding.

Along a flux surface, the perturbation spreads much faster than it does across surfaces. We follow this spreading on the flux surface of the deposition peak ($\psi$ = 0.9463). Once the injection starts, the pressure gradient of the source region drives two counter-propagating flows along the field lines, with opposite signs of the velocity on the two sides of the injection point (Fig. 10(a)) The near-source region is quasi-stationary during the injection. The sign reversal of the velocity profile propagates outward at a parallel velocity of about 0.75 $c_s$. A pressure-driven flow comparable to $c_s$ is expected. The two reversals propagate outward on their respective sides and meet on the opposite side about 0.1 ms after the injection begins ($t \approx 1.11$ ms). The peak mass flow velocity on the two sides of the source region is only about 0.016 $c_s$, which shows that the injection is a weak perturbation to the background. The flow structure and the mass transport proceed on clearly different time scales. After the injection stops ($t$ > 2 ms) the source disappears, the velocity decays, and it is close to zero by $t \approx 6$ ms (Fig. 10(b)). the ion density increment is concentrated at the injection point during the injection (Fig. 10(c)), then decreases monotonically towards the opposite side, and becomes poloidally uniform within about 1 ms after the injection stops, with no clear poloidal structure. the change of electron temperature reaches its maximum during the injection and recovers afterwards, and is nearly poloidally uniform (Fig. 10(d)). This reflects the much faster energy transport compared with the particle transport. Finally, the high ion velocities appear only in a narrow radial band near the outer boundary and do not form a large-scale shear flow which could affect the global transport.

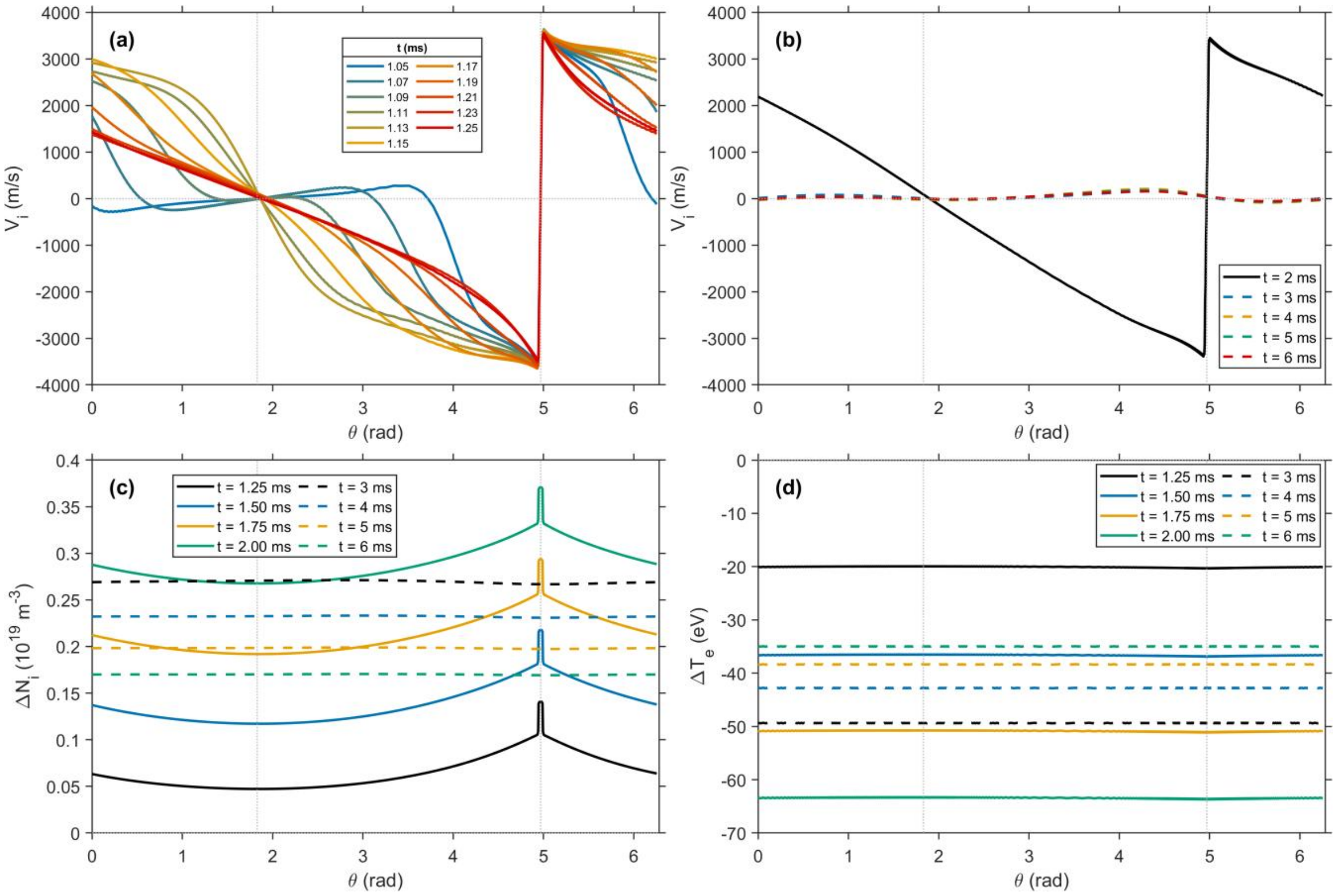

**Fig. 10.** Poloidal profiles on the flux surface of the deposition peak ($\psi$ = 0.9463, $R$ = 2.2782 m). The abscissa is the poloidal angle $\theta$, and the dotted lines mark the injection location ($\theta$ = 4.970 rad) and its opposite side (1.829 rad). (a) Parallel ion velocity $V_i$ during the injection ($t$ = 1.05–1.25 ms in steps of 0.02 ms, blue to red); (b) $V_i$ after the injection ($t$ = 2/3/4/5/6 ms, solid and dashed); (c) ion density increment $\Delta N_i$ relative to $t$ = 1 ms, which at $t$ = 1.25 ms is $0.141 \times 10^{19}$ m$^{-3}$ at the injection point and $0.047 \times 10^{19}$ m$^{-3}$ on the opposite side; (d) electron temperature increment $\Delta T_e$ on the same baseline, with a minimum of −63.7 eV during the injection and −35.0 eV at the end of the pulse. The poloidal circumference of this flux surface is 3.417 m.

## 5.2 Validations with experiment

The comparison with experiment needs a well-defined basis. As mentioned in Section 4.2, The simulated injected flux is fixed by the calibration, so the criteria for the comparison with experiment are dimensionless ratios and spatial positions.

The first comparison is the deposition position (Fig.11). The x-axis is the Gaussian gradient coordinate $(R - R_c)/\sigma$, in which $R_c$ and $\sigma$ are the center and width of a Gaussian fit to the gradient of the respective background density profile. We Take the characteristics of the background profiles rather than the deposition peaks as the reference to avoid circular reasoning.. The deposition peaks of the three simulated cases lie between −0.10 and −0.24 $\sigma$, all of them inside the plateau at the top of the experimental peak ($[-1.10, 0.26]\,\sigma$). The deposition increment is about 19% of the pedestal-top density. The peak positions of the three cases differ by no more than 0.8 mm. as $V_m$ decreases across the scanned range, the peak height changes while the position and width do not. Taken together, the deposition occurs in the mid-pedestal region and its position agrees with the experiment in the relative pedestal coordinate.

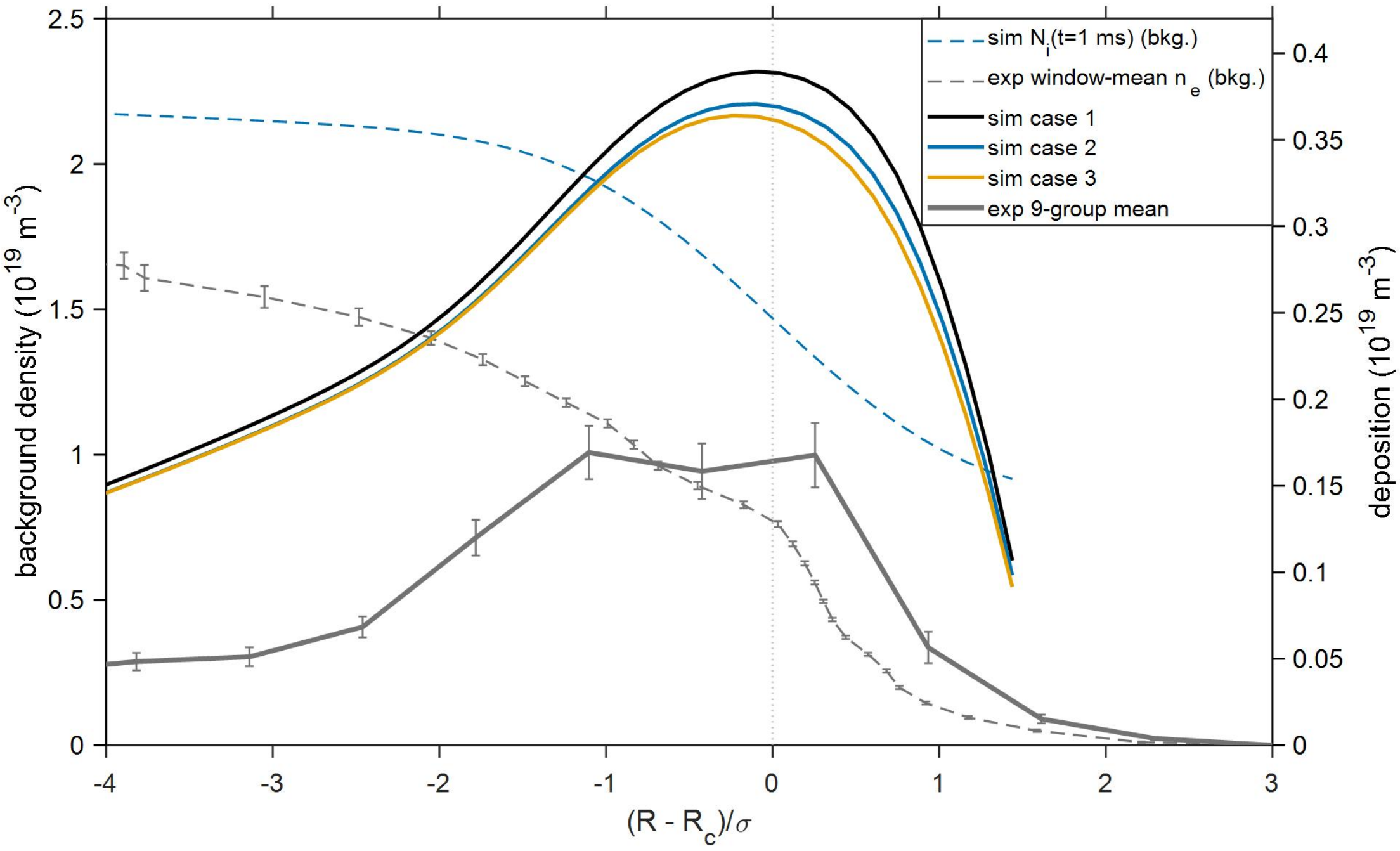

**Fig. 11.** Comparison of the simulated and experimental deposition profiles, the abscissa being the Gaussian fit coordinate of the gradient of the respective background density profile (for the simulation, fitted to $|dN_i/dR|$ at $t$ = 1 ms, giving $R_c$ = 2.2787 m and $\sigma$ = 5.35 mm. For the experiment, fitted to the window-averaged $n_e$, giving $R_c$ = 2.3013 m and $\sigma$ = 13.35 mm). Left axis, background density profiles (simulated $N_i$ at $t$ = 1 ms, blue dashed; window-averaged measured $n_e$, gray dashed). Right axis, deposition increments (three cases, $\Delta N_i$ at $t$ = 2 ms, black, blue and orange solid lines with peak heights of $0.389 \times 10^{19}$ m$^{-3}$, $0.371 \times 10^{19}$ m$^{-3}$ and $0.364 \times 10^{19}$ m$^{-3}$; nine-group mean measured source profile, thick gray solid line). The vertical bars on the measured profiles show their uncertainties: the standard deviation of the measured density within the analysis window (1.98–2.50 s) for the background profile, and the standard error of the mean over the nine pulse groups for the measured source profile.The vertical dotted lines mark the respective pedestal gradient centers. The two data sets are taken from different stages of the discharge and the overall levels of the background profiles differ; the absolute amplitudes are not used as a validation criterion. The experimental profiles span R = 1.92–2.35 m and the simulated profiles R = 2.19–2.29 m. Only the part near pedestal is shown.

The second test is how much of the injected particle content remains. Table 5 lists the retention, defined as $\Delta N_i(t)/2\Sigma S_{\text{dissp}}$ . the denominator is the total dissociation in ion equivalents, and the integration is taken over the whole simulation domain. The dissociation of the three cases amounts to 95–101% of the theoretical injection, which shows that the denominator is reliable. Moreover, $2\Sigma S_{\text{dissp}} + \Sigma S_{\text{recp}}$ agrees with $\Sigma S_{Ip}$ to within $\pm$0.5% in all three cases. Therefore, all the dissociation products enter the ion channel and the accounting is self-consistent. The simulated conversion at the end of the injection is 92.1–92.5%, and the retention at the end of the pulse is 77.2–78.1%, against experimental values of 94.2/94.3% and 71.2/71.4%, respectively. The simulation thus lies ~2% below the experiment at the end of the injection , and ~6% above it at the end of the pulse. this reversal between the end of the injection and the end of the pulse shows that the perturbation decays more slowly in the simulation than in the experiment. The simulated retention falls by ~14.7% from the end of the injection to the end of the pulse against ~23.0% in the experiment.

Both validation quantities are robust over the scanned injection velocity (300–1000 m/s). The retention of the three cases scatters by no more than ~0.5% at the end of injection (1 ms) and by no more than ~1% at the end of pulse (5 ms) and the deposition peak positions differ by no more than 0.8 mm. When the boundary condition of the ion density at the outer boundary is replaced by Neumann, the way the perturbation accumulates at the edge changes (the deposition peak moves outward and the edge becomes a net source), but the validation result concerning the deposition position in the pedestal is unchanged. Therefore, this setting is used here as a boundary sensitivity control only. Possible explanations of the remaining differences, including the absorption of particles by the outer boundary condition, the absence of the SOL, the omission of turbulent transport and the temperature dependence of the rate coefficients, are discussed in Section 6.

| | $R$(1 ms) (%) | $R$(5 ms) (%) |
|---|---|---|
| case 1 | 92.1 | 77.2 |
| case 2 | 92.3 | 77.6 |
| case 3 | 92.5 | 78.1 |
| experiment | 94.2 / 94.3 | 71.2 / 71.4 |

**Table 5.** Ion retention in the simulations and in the experiment. $R$(1 ms) is the fraction of the dissociated molecules converted into the ion inventory by the end of the injection (simulation time $t$ = 2 ms) and $R$(5 ms) is the fraction remaining at the end of the pulse ($t$ = 6 ms). The simulated values are $\Delta N_i(t)/2\Sigma S_{\mathrm{dissp}}$ integrated over the domain; the experimental values are the ratio of the net increment to the gross injected amount (net/gross), the two columns being the whole-volume and core values.

## 6. Discussion

The simulated and experimental retention differ in two respects. In the simulation, during the injection, part of the newly ionized population is created close to the outer boundary, where the density is held fixed by the boundary condition. These ions are removed before they can be transported inward, which lowers the conversion relative to the experiment. After the injection, the perturbation decays more slowly than observed, because the channels that carry it away in the experiment are not represented. These channels are the parallel losses along open field lines to the divertor, the turbulence excited by the perturbation itself, and the density feedback control of the discharge. They are also among the processes that maintain the background in the discharge. The quantitative contribution of each channel is beyond the capability of the present model. The coefficients used here were, moreover, constructed to maintain the background rather than to describe its response. In these simulations the injected ions can be lost only through the boundaries or by recombination, and the retained fraction reaches about 100% at the end of the pulse when the outer boundary condition is replaced by Neumann. the loss is therefore carried by the boundary flow, with internal sinks such as recombination playing only a minor part.

Both comparisons are also affected by the state of the discharge. The equilibrium and initial profiles of the simulation are taken from a later flat-top stage of shot 71438 ($t$ = 5.059 s), whereas refers to the density ramp-up phase. The background density profiles therefore differ in their overall level (Fig. 11), and only relative quantities are compared. The experimental profiles are used as measured, on the axis of the measured channels, whereas the simulated equilibrium is an kinetic-EFIT reconstruction of the later stage. The radial position of the simulation lies about 2 cm inside that of the experiment, sothe particle increment inside the separatrix inferred from the experimental data carries a systematic uncertainty of that scale , together with the overestimation of the SOL density introduced by the profile mapping onto the flux surfaces (discussed below).

The comparison of the deposition position of the simulations in Ref. [8,9] is more direct. The referred deposition region moves inward while the boundary temperature falls to the 10 eV level, at which the dissociation rate decreases steeply and the molecules survive further inside. Neither element occurs in the present simulations for three reasons. First, the discharges differ. In Ref. [8] the penetration depth is larger in L-mode, where the lower density and temperature reduce the dissociation rate along the beam path. By contrast, the equilibrium analyzed here is an H-mode discharge whose electron temperature at the dissociation location stays at the 230–300 eV level throughout the pulse. The molecules are therefore dissociated within the outermost cells, and the deposition peak is built by the inward transport of the resulting atoms and ions rather than by a motion of the source. Second, the injected flux is lower, $3 \times 10^{21}\ \mathrm{m^{-2}\,s^{-1}}$ here against about $10^{22}\ \mathrm{m^{-2}\,s^{-1}}$ in Ref. [8] and $8 \times 10^{21}\ \mathrm{m^{-2}\,s^{-1}}$ in Ref. [9]. A larger injected amount deepens the penetration only through the cooling it produces. Because the molecules dilute the plasma along the beam path, and the dissociation rate falls once the temperature drop is large enough, moving the deposition inward. Here the flux is low and the edge temperature stays on the plateau of the rate coefficients, so this path is not open either. Third, the peak is maintained by the balance between the local dissociation source and the perpendicular transport. Since $S_{Ip} \propto N_n N_i$, the source increases with the local density, and as long as the profiles are not strongly redistributed, as in these simulations, the peak remains where it is.

The transport coefficients are held fixed during the pulse, so that the response of the transport to the fueling perturbation is not described. Responses of this kind, such as a prompt increase of density fluctuations and a coherent mode after SMBI in ohmic plasmas on EAST [25,26], or ELM crashes relaxing the pedestal on a few-millisecond time scale at ASDEX Upgrade [24], may be present in the discharge analyzed in Section 3, and their contribution to the background evolution during the window is not quantified here.

A further limitation is the reduced form of the electron energy equation adopted in Ref. [8], in which the electron temperature is set by parallel heat conduction, perpendicular diffusion and the local atomic processes. Consistent with this reduction, the electron flow is not an independent degree of freedom, and its advection and compression terms are absent from the equation. The advective term is negligible here, the electron temperature varying by less than 0.3% along a field line. This flatness is not an artifact of an excessively large parallel conduction: the parallel thermal conductivity is limited by the free-streaming expression (Appendix 1), so that without the limiter it would be larger and the electron temperature flatter still, and the neglect of the advective term is conservative. The compression term is not. Its strength can be estimated from the compression term of the ion equation in Table 4. With $\nabla \cdot \boldsymbol{V}_e \approx \nabla \cdot \boldsymbol{V}_i$, its effect on the electron temperature is similar to the ion temperature. From the compression term and the change of $T_i$ in Table 4, it would lower $T_e$ by about 40 eV with respect to the case without this term. The simulated electron temperature (180–824 eV) remains well above the 50 eV at which the dissociation and ionization rates begin to fall steeply, so the omission does not affect the conclusion. A self-consistent inclusion of the term would require an additional closure for the electron flow, outside the scope of the model used here.

The temperature equations contain no explicit term for the internal energy carried by the particle flux, $(3/2)T\Gamma$. The particle transport moves density without carrying energy, so the injected particles affect the temperature profiles only through the local atomic processes. In earlier applications [20] this term was represented by an additional correction to the $\chi$ coefficients. Here it is not represented explicitly. Its effect, like that of the other processes omitted from the model, is contained in the effective coefficients described in Section 4.1. The three transport channels are also constructed independently, with $\chi_i$, $\chi_e$ and $D_\perp$ obtained from the experimental $T_i$, $T_e$ and $N_i$ profiles, so the relative behavior of the profiles, for instance $T_i/T_e$, is not constrained by the construction, whose outcome is the shape of the coefficients only.

The simulated domain imposes two further limitations. It ends at the outermost flux surface considered, so no SOL or divertor region is included and the parallel losses that remove particles and energy from the edge of a real discharge enter only through the boundary conditions. Both Dirichlet and Neumann conditions are idealizations for the plasma–wall interface. A self-consistent treatment of either would require coupling to an edge transport code, and the comparison between the two in Section 5.2 is therefore a sensitivity measure rather than a physical result. A separate limitation concerns the experimental analysis. There the measured one-dimensional profile is mapped onto flux surfaces assuming poloidal symmetry. While in the SOL and divertors, the actual transport makes the density poloidally asymmetric and intrinsically lower, which overestimates the local density and hence the particle increment inferred from the experiment.

The dissociation and ionization rates are evaluated at the local electron temperature of the simulation, which is 230–300 eV at the dissociation location and therefore on the high-temperature plateau of the fitted coefficients. the steep dependence that sets in below about 50 eV is not resolved, and the regime in which a cooling edge deepens the penetration, as in Ref. [8,9], is not reached. The molecules are assumed to lose their directed velocity upon dissociation, their kinetic energy being small compared with both the internal energy of the background and the energies of the atomic processes. Finally, the analysis assumes a separation of time scales: only the evolution correlated with the pulse train is retained by the group-by-group fitting, while background changes slower than one pulse group are removed as a smooth trend, and fluctuations faster than the fit windows are averaged out.

Notwithstanding these limitations, the way the source is treated determines what the comparison can establish. Because the source is calibrated by the experimentally inferred increment and no further parameter is adjusted afterwards, the agreement of the deposition peak with the experimental one tests the transport model itself. In previous comparisons the source intensity was prescribed independently of the experimental density response, so the comparison tested the profile shape rather than the particle balance. here it is fixed by the experimental increment before the comparison is made. The conservation requirement and the separation between the shape of the coefficients and their calibrated level are properties of the transport coefficients constructed here rather than SMBI, and apply to any flux-coordinate transport code in which the coefficients are prescribed from any profiles.

Of the limitations above, those acting most directly on the present results are the frozen transport coefficients and the reduced electron energy equation. The next step is to let the transport respond to the perturbation, either by coupling to a turbulence code or by inferring the coefficients as functions of time from the measured evolution, with the electron flow closure introduced at the same stage. Fast events such as ELM crashes would then have to be treated explicitly rather than through their possible presence within the analyzed window.

## 7. Conclusions

This work has established a quantitative link between SMBI fueling simulations and experiment through the particle increment per pulse. On the experimental side, the density evolution measured by microwave reflectometry in EAST shot 71438 was processed with a piecewise fitting procedure applied group by group, which separates the SMBI-induced density increase from the background evolution and from uncorrelated perturbations. The decay-corrected increase rate peaks in the mid-pedestal region for all nine pulse groups, and its volume integration gives the particle increment per pulse, $0.424 \times 10^{19}$ molecules over the measurable volume and $0.240 \times 10^{19}$ within the core, with 94.2% (whole volume) and 94.3% (core) of the increment retained at the end of the injection, and 71.2% and 71.4%, respectively, at the end of the pulse. These values provide the experimental reference for the simulations.

On the simulation side, two properties of the trans-neut code were put in place before the comparison was made. The poloidal advection operator of the ion equation and the perpendicular diffusion operator of the atom equation were recast into conservative flux form, reducing the closure residual of the particle budget to about 2% of the injected amount. The perpendicular transport coefficients were constructed from the discrete radial flux balance with the local metric factor retained cell by cell, reducing the unphysical changes of the initial density and temperature profiles in source-free evolution by about an order of magnitude relative to the conventional continuous inversion. Both elements are independent of the SMBI problem itself and apply to any flux-coordinate transport code in which the coefficients are prescribed from experimental profiles.

Three cases sharing the same nominal molecular flux but different combinations of molecular density and velocity were simulated, with the source calibrated by the experimental increment. The simulated deposition peak lies in the mid-pedestal region and agrees with the experimental peak in the relative pedestal coordinate, with the peak positions of the three cases nearly coincident. The simulated conversion at the end of the injection, 92.1–92.5%, lies slightly below the experimental value 94.2%/94.3% (whole volume/core), while the retention at the end of the pulse, 77.2–78.1%, lies above the experimental 71.2%/71.4%. The perturbation thus decays more slowly in the simulation, which is attributed to the loss channels not represented in the model.

The simulations further show how the deposition is built. The injected molecules dissociate within the outermost cells, and the deposition peak in the pedestal is built by the inward transport of the resulting atoms and ions rather than by a motion of the source. The density rise is rapidly spread over the flux surface by fast parallel convection, becoming poloidally uniform within about 1 ms after the injection stops, while the radial transport remains diffusive and slow. The edge electron temperature stays on the high-temperature plateau of the dissociation and ionization rate coefficients throughout the pulse, so that the dissociation remains localized at the edge, and the inward shift of the deposition with decreasing temperature reported in earlier simulations does not occur under the present conditions.

The main limitations of the present model are the frozen transport coefficients, which cannot respond to the fueling perturbation, and the reduced electron energy equation, which omits the compression work of the electron flow. The next step is to let the transport coefficients evolve with the perturbation, either by coupling to a turbulence model or by inferring them as functions of time from the measured evolution, together with a self-consistent closure of the electron flow. Fast events such as ELM crashes would then be treated explicitly.

# Appendix 1. Model

## A1.1 Model equations

The simulation code used in this work is a modified version of the trans-neut code developed by Wang et al. [8], which is built on the BOUT++ framework [19] and has been applied to SMBI simulations in earlier studies [8–11]. The model follows four elementary processes, namely the dissociation of the injected molecules into atoms, the ionization of atoms into ions, the recombination of ions with electrons into atoms, and charge exchange between ions and atoms. The molecular beam is treated as a fluid and is transported along the injection direction, whereas ions and atoms are transported along and across the magnetic field. The equations fall into two categories, the transport equations of the three species and the particle-interaction terms through which mass and energy are exchanged between them. With the quasi-neutrality assumption $N_e = N_i$, no separate electron density equation is required.

The transport equations of the ions, atoms, and molecules, together with the definitions of the four source terms, are given as Eqs. (1)–(5) and are not repeated here. The equations that complete the model are those of the electron and ion temperatures and of the ion velocity,

$$\frac{\partial T_e}{\partial t} = \frac{2}{3}\frac{1}{JN_i}\frac{\partial}{\partial y}\left(\frac{J\kappa_{\parallel e}}{g_{22}}\frac{\partial T_e}{\partial y}\right) + \frac{2}{3}\nabla_\perp\cdot(\chi_{\perp e}\nabla_\perp T_e) - \nu_{\text{ionz}}\left(T_e + \frac{2}{3}W_{\text{ionz}}\right) - \frac{2}{3}\nu_{\text{diss}}(W_{\text{diss}} + W_{\text{bind}}) - 2\frac{m_e}{M_i}\frac{T_e - T_i}{\tau_{ei}} + \nu_{\text{rec}}W_{\text{rec}} \qquad (6)$$

$$\frac{\partial T_i}{\partial t} = -V_{\parallel i}\partial_\parallel T_i - \frac{2}{3}\frac{T_i}{J}\frac{\partial}{\partial y}\left(\frac{JV_{\parallel i}}{\sqrt{g_{22}}}\right) + \frac{2}{3}\frac{1}{JN_i}\frac{\partial}{\partial y}\left(\frac{J\kappa_{\parallel i}}{g_{22}}\frac{\partial T_i}{\partial y}\right) + \frac{2}{3}\nabla_\perp\cdot(\chi_{\perp i}\nabla_\perp T_i) - \nu_{\text{ionz}}T_i - \frac{2}{3}\nu_{CX}(T_i - T_n) + 2\frac{m_e}{M_i}\frac{T_e - T_i}{\tau_{ei}} + \nu_{\text{rec}}T_i \qquad (7)$$

$$\frac{\partial V_{\parallel i}}{\partial t} = -V_{\parallel i}\partial_\parallel V_{\parallel i} - \frac{1}{N_iM_i}\partial_\parallel p_{ei} + \frac{4}{3}\frac{1}{JN_iM_i}\frac{\partial}{\partial y}\left(\frac{J\eta_{0i}}{g_{22}}\frac{\partial V_{\parallel i}}{\partial y}\right) - (\nu_{\text{ionz}} + \nu_{CX})V_{\parallel i} \qquad (8)$$

where $\partial_\parallel = (1/\sqrt{g_{22}})\,\partial_y$ is the parallel derivative, $p_{ei} = N_i(T_e + T_i)$ is the combined electron and ion pressure, $T_n$ is the neutral temperature ($T_n \equiv T_i$ in the model), $T_m = 300$ K is the molecular temperature of the injected material (room temperature), at which the ionization dilutes $T_e$ and $T_i$, $T_m \ll T_i$, $W_{\text{ionz}} = 20$ eV, $W_{\text{diss}} + W_{\text{bind}} = 5$ eV and $W_{\text{rec}} = 4.5$ eV are the energies involved in the ionization, dissociation, and recombination reactions, and the reaction and collision frequencies are defined below. The neutral parallel velocity is not evolved, $V_{\parallel n} = 0$, the parallel transport of atoms being diffusive. The electron flow is likewise not treated as an independent degree of freedom, so that its advection and compression terms do not appear in Eq. (6).

The parallel transport coefficients are the classical ones. The parallel thermal conductivities are $\kappa_{\parallel e} = 3.2\,N_eT_e\tau_e/m_e$ and $\kappa_{\parallel i} = 3.9\,N_iT_i\tau_i/m_i$, the parallel ion viscosity is $\eta_{0i} = 0.96\,N_iT_i\tau_i$, and the electron-ion energy exchange time is $\tau_{ei} = 1/\nu_{ei}$ with

$$\nu_{ei} = 2.91\times10^{-6}\frac{N_e\lambda_{ei}}{T_{e,\text{eV}}^{3/2}}\ \text{s}^{-1}, \qquad \lambda_{ei} = 24 - \ln\left(\frac{\sqrt{N_e/\text{cm}^{-3}}}{T_{e,\text{eV}}}\right) \qquad (9)$$

with $N_e$ in $\text{cm}^{-3}$ and $T_e$ in eV. $\lambda_{ei}$ is evaluated once from the reference density and temperature and held fixed, while $\tau_{ei}$ is updated with time. Both conductivities are limited by a free-streaming expression, following the earlier applications [8] in both the form of the limiter and the choice of its parameters. The limiting value is $\kappa_{\text{fl}} = q_{95} L_0 N V_{th}$, where $N$ is the density of the respective species, $V_{th} = (T/m)^{1/2}$ is the thermal velocity, $L_0$ the normalizing length of the simulation (2.29 m here) and $q_{95} = 5$ the safety factor at $\psi = 0.95$. The coefficient actually used is $\kappa_{\text{fl}}\kappa_0/(\kappa_{\text{fl}} + \kappa_0)$, with $\kappa_0$ the classical value above. In these simulations the classical conductivities lie far above the limiting value, so that the classical conduction cannot be applied without limiting it.

The four reaction-rate coefficients are evaluated with empirical fits. For ionization and for dissociation,

$$\langle\sigma v\rangle_{\text{ionz}} = \langle\sigma v\rangle_{\text{diss}} = 3 \times 10^{-8} \frac{(0.1T_e)^2}{3 + (0.1T_e)^2} \ \text{cm}^3\, s^{-1},$$

$$\langle\sigma v\rangle_{CX} = \left[1.7 + 1.9\, \frac{(1.5T_i)^{1/3} - 15^{1/3}}{(150T_i)^{1/3} - 15^{1/3}}\right] \times 10^{-8} \ \text{cm}^3\, s^{-1} \qquad (10)$$

with the temperatures in eV, and for recombination the radiative-recombination fit of Seaton [27],

$$\langle\sigma v\rangle_{\text{rec}} = 5.197 \times 10^{-14}\, Z\sqrt{\Lambda}\left(0.4288 + 0.5\ln\Lambda + \frac{0.469}{\Lambda^{1/3}}\right) \ \text{cm}^3\, s^{-1}, \qquad \Lambda = \frac{13.6\ \text{eV}}{T_e} \qquad (11)$$

In Eq. (10) the same fit is used for the dissociation and ionization rates, as in the earlier applications [8–11]. Fig. 12 shows the ionization coefficient of Eq. (10) together with the measured values on which it is based [28].

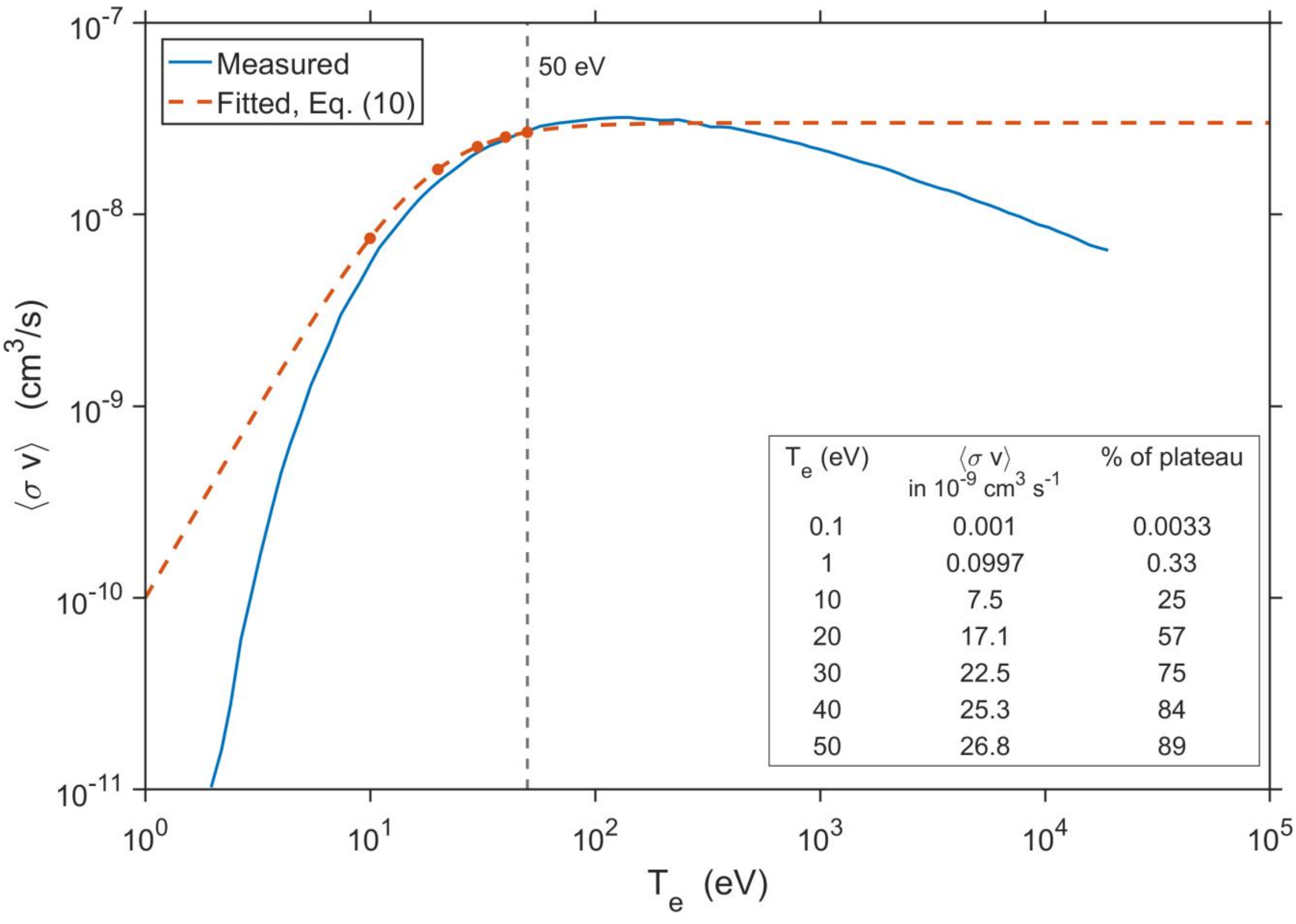

**Fig. 12.** Ionization rate coefficient $\langle\sigma v\rangle$ as a function of the electron temperature. Solid curve, measured values [28]; dashed curve, the fit of Eq. (10) used in the simulations. The vertical dashed line marks 50 eV, above which the coefficient is nearly constant. The values of the fit at 0.1, 1, 10, 20, 30, 40 and 50 eV are listed in the figure together with their ratio to the plateau value, and the last five are marked on the curve.

The transport operators of Eqs. (1)–(8) are evaluated in the conservative flux forms described in Section 4.3, which differ from those used in the earlier applications [8–11]. Two of them are written out here, and the others follow the same pattern. The parallel advection term of Eq. (1) is evaluated as

$$-\frac{1}{J}\frac{\partial}{\partial y}\left(\frac{JV_{\parallel i}N_i}{\sqrt{g_{22}}}\right) = -\frac{\sqrt{g_{22}}}{J}\left[W\,\partial_\parallel V_{\parallel i} + V_{\parallel i}\,\partial_\parallel W\right], \qquad W = \frac{JN_i}{\sqrt{g_{22}}} \qquad (12)$$

and the perpendicular diffusion term of Eq. (2) as

$$D_\perp g^{11}\frac{\partial^2 N_n}{\partial x^2} + \frac{1}{J}\frac{\partial}{\partial x}\left(JD_\perp g^{11}\right)\frac{\partial N_n}{\partial x} \qquad (13)$$

that is, the flux form $\nabla_\perp \cdot (D_\perp \nabla_\perp N_n)$ with the metric factor kept inside the derivative of the coefficient and the second derivative evaluated directly on the radial grid. Written in this way the poloidally summed particle number is conserved by construction, whereas the forms used previously, $V_{\parallel i}\partial_\parallel N_i$ for the advection and $D_\perp \nabla_\perp^2 N_n$ for the diffusion, are not. The remaining modifications are the conservative flux forms of the parallel heat conduction in Eqs. (6) and (7), of the compression term of Eq. (7), of the parallel viscosity in Eq. (8) and of the parallel atom diffusion in Eq. (2). Finally, the perpendicular transport coefficients are constructed inside the code from the initial profiles instead of being read from the grid file, as described in A1.2.

## A1.2 Construction of the perpendicular transport coefficients

Inferring the perpendicular transport coefficients by inverting the steady-state transport equations with measured profiles is an established practice in BOUT++ transport modeling [20,22,23]. The coefficients are interpreted as an effective representation of the processes that sustain the H-mode background but are not modeled explicitly, and are held fixed during the pulse. Here they are constructed inside the code from the initial profiles rather than read from the grid file, so that the model contains no core particle or heat source and the coefficients are not free parameters.

### *(a) Conventional method*

In the conventional approach the inversion is performed on the continuous, steady-state equations, without drifts or volume sources,

$$\nabla_\perp \cdot (D_\perp \nabla_\perp N) = 0 \quad \Rightarrow \quad D_\perp \frac{\partial N}{\partial x} = -\Gamma|_{\mathrm{core}}, \qquad N\chi_\perp \frac{\partial T}{\partial x} = -Q|_{\mathrm{core}}$$

that is, the radial flux is required to be uniform, its value being fixed by the core (inner-boundary) condition. In the temperature channel the inversion also retains the correction for the internal energy carried by the particle flux, which the earlier applications folded into the coefficients [20], that correction is not part of the construction used here. The resulting coefficients are computed offline and passed to the discrete solver as frozen constants. Two features of this procedure are relevant below. The grid geometry does not enter the inversion, and the outcome is a single radial profile, that is, the coefficients are poloidally uniform.

***(b) Method used here***

The requirement is the same as in the conventional construction. The two differ only in the object on which it is imposed. Here we require the discrete face flux of the operator that the code actually integrates,

$$F_f = (Jg^{11})_f\,\chi_f\,\frac{T_{i+1} - T_i}{\Delta x_f}$$

to be uniform on every face, $F_f = F_0$. Solving face by face gives

$$\chi_f = \frac{F_0}{(Jg^{11})_f\,(\partial_x T)_f}, \qquad (Jg^{11})_f = 1/2\,[(Jg^{11})_i + (Jg^{11})_{i+1}]$$

so that the metric factor $Jg^{11}$ enters the coefficients cell by cell. At the injection trajectory it increases by about 40% from the inner boundary to the edge, and along a flux surface it varies by a factor of about 2.4. The coefficient is required at cell centers, whereas the relation above defines it on faces. The two are linked by the face average $\chi_{f,i} = 1/2\,(\chi_i + \chi_{i+1})$, whose exact inverse, $\chi_{i+1} = 2\chi_{f,i} - \chi_i$, is used so that the cell-to-face averaging reproduces the prescribed face values exactly. The inverse leaves the alternating component of the cell-center field undetermined, and this is fixed by choosing, among the fields that reproduce the same face values, the smoothest one, which for the density channel is also the one that keeps the coefficients positive. This is the form used in the simulations reported here, where the face values remain exact for all three channels. Recomputing the face fluxes with the coefficients as written into the field gives a flux-constancy residual $\max|F - \bar{F}|/|\bar{F}|$ at machine precision for the three channels. A simple face average is retained as a fallback for sequences in which no positive cell-center field reproduces the face values. The construction is carried out separately for every poloidal row, each using its own $(\psi, J, g^{11})$, so the coefficients are two-dimensional, $\chi(x, y)$. As in the earlier practice, the coefficients are normalized to 1 $m^2\,s^{-1}$ at the inner domain boundary ($\psi \simeq 0.55$, not at the magnetic axis). The inversion therefore fixes their radial and poloidal shape, while their absolute level remains a calibrated, rather than first-principles, quantity. In a geometry with a uniform metric factor, and on a grid fine enough for the discrete and continuous fluxes to coincide, this construction reduces to the conventional one.

Fig. 13 shows the resulting coefficients for the electron channel.

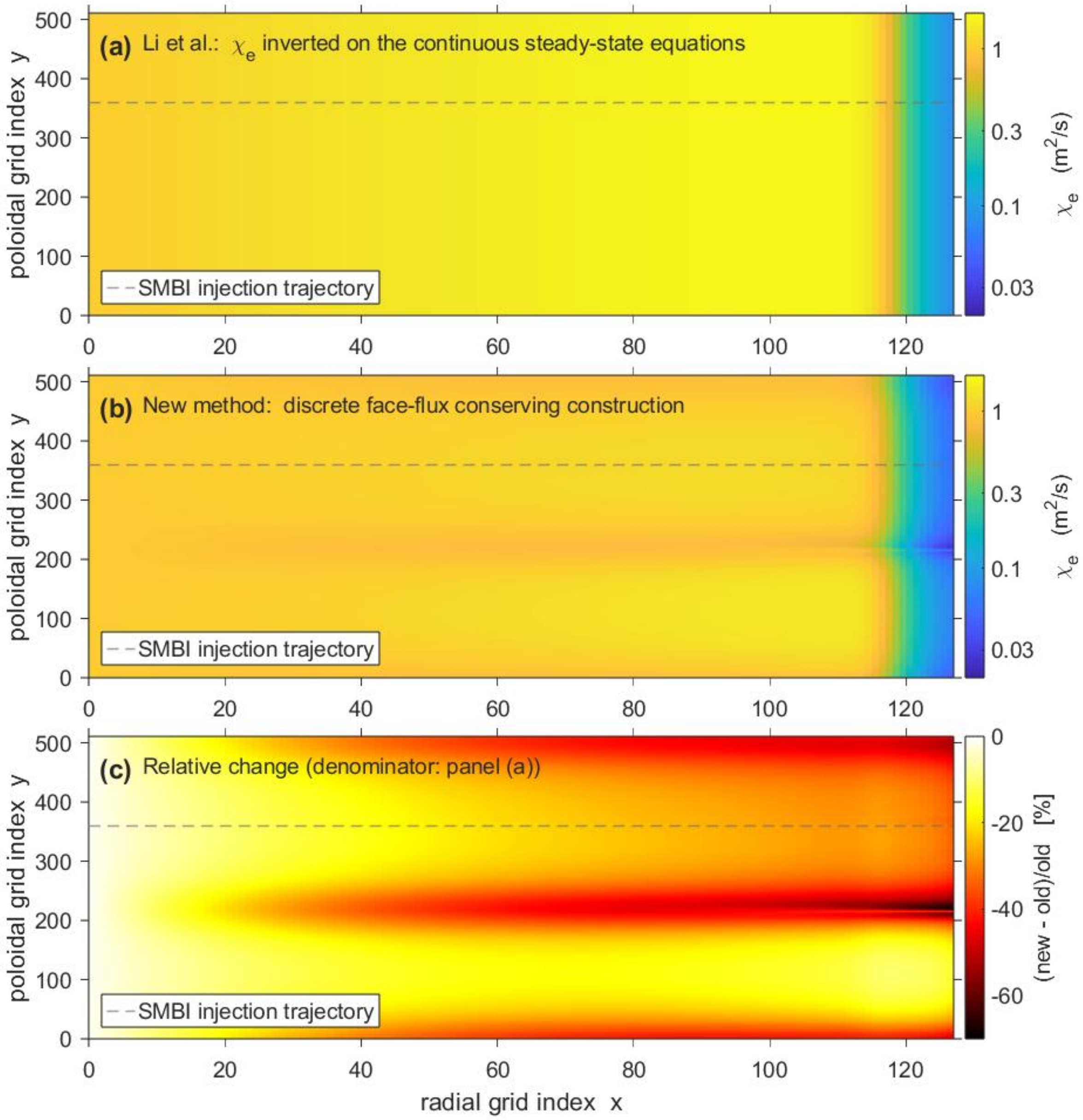

**Fig. 13.** Perpendicular electron thermal diffusivity $\chi_e$ over the full $(x, y)$ grid. (a) Conventional values, obtained by the method of Li et al. [20], that is, by inverting the continuous steady-state equations with measured profiles, and read from the grid file as frozen coefficients. (b) Values constructed here from the discrete face-flux balance. (c) Relative change, (new − conventional)/conventional, taking the conventional values as the denominator; the values span −69% to −1%, with a median of −22%. The dashed line marks the SMBI injection trajectory ($y$ = 359). The conventional coefficients are poloidally uniform, a single value at each radial position, whereas the new construction varies cell by cell with the local metric factor; the largest reductions occur where the poloidal variation of $Jg^{11}$ is strongest.

As a corroboration, the source-free stage should leave the profiles unchanged. The three profiles do remain approximately constant, and the comparison of the two constructions over that stage is given in Table 2.

## Appendix 2. Coordinates and geometry

The injection parameters of the code are the molecular density and velocity, prescribed in a narrow strip near the OMP (Fig. 6). Since the beam propagates almost perpendicularly to the magnetic field, the strip is kept narrow so that the molecular trajectories can be treated as rectilinear along the injection direction.

Simulations are performed on a 2.5D flux-coordinate grid with 128 radial and 512 poloidal points (132 radial points including the guard cells) and one toroidal plane, covering the normalized poloidal flux range $\psi \in [0.55, 0.99]$ (Fig. 6, where every sixteenth grid line is drawn). The domain is limited to the confined region. The region beyond the last closed flux surface is not included, since its complex magnetic geometry would substantially increase the computational cost and since particles crossing the last closed flux surface are assumed not to return. It enters the simulations only through the boundary conditions of Appendix 3. The inner cutoff at $\psi = 0.55$ balances two effects. The injection also raises the ion density at $\psi < 0.55$, so that a deeper cutoff would capture more of the perturbation, but the transport mechanisms that the model does not describe become more pronounced closer to the core, so that a deeper domain would increasingly be governed by physics outside the model.

The relatively high poloidal resolution is required by the particle-number conservation of the scheme. The finite-difference discretization used here conserves particle number only approximately, and the conservation error decreases with the grid resolution. The radial extent of the domain is only about 10 cm, much smaller than the poloidal circumference (3.4 m at the deposition surface, Fig. 10), so that an acceptable conservation error requires a correspondingly fine poloidal grid.

All the data used in the simulations are taken from EAST shot 71438. The magnetic equilibrium is reconstructed from the experimental data with EFIT, and the density and temperature profiles are mapped onto the resulting flux coordinates.

# Appendix 3. Boundary conditions

The boundary conditions are chosen according to three principles, namely consistency with the physical model, maintenance of the profiles in the absence of the injection, and the use of Neumann conditions by default where no specific requirement applies. Table 6 lists the conditions used. A Dirichlet condition fixes the value of the corresponding profile at the boundary, and a Neumann condition its gradient there. The values are chosen so that the initial profiles are maintained in the absence of the injection, and the region of influence of the conditions is shown in Fig. 6.

| Variable | Inner boundary | Outer boundary |
| --- | --- | --- |
| $N_i$ | Neumann | Dirichlet |
| $T_e$ | Dirichlet | Neumann |
| $T_i$ | Dirichlet | Dirichlet |
| $V_{\parallel i}$ | Neumann | Neumann |
| $N_n$, $N_m$, $V_m$ | Neumann | Neumann |

**Table 6.** Boundary conditions of the simulations.

The neutral temperature and the neutral parallel velocity do not evolve ($T_n \equiv T_i$, $V_{\parallel n}$ = 0) and their conditions are Neumann.

For the ion density, a Neumann condition is applied at the inner boundary so that the inward transport of the injected particles is not impeded, and a Dirichlet condition at the outer boundary for numerical stability. Holding the outer density at its initial value has two consequences, both of which appear in the results. Part of the newly ionized population is created close to the outer boundary and is removed there before it can be transported inward, which lowers the fraction of the injected molecules that is retained as ions. And the density increment peaks slightly inside the boundary rather than at it, with a lower density in the outermost cells.

For the electron and ion temperatures, the Dirichlet conditions at the inner boundary follow from the observation, confirmed by both the experiments and the simulations, that the injection has little effect on the temperature at that location. At the outer boundary the two are treated differently. The ion temperature retains the Dirichlet condition of the earlier applications [8], where such boundary values represent the effective local sources and fluxes that maintain the profiles at steady state. The electron temperature is instead given a Neumann condition, unlike in Ref. [8], because the cooling of the edge region controls how far the molecules penetrate, and fixing its value there would suppress that cooling, whose effect on the penetration depth is one of the questions examined here.

The penetration depth of the beam is sensitive to the boundary temperature. The reaction-rate coefficients of Eq. (10) fall steeply below about 50 eV (Fig. 12), so that a boundary temperature of that order would considerably increase the survival of the molecules and allow them to penetrate further inward, as reported in Ref. [8,9]. To obtain an upper bound on the cooling that the injection could produce, we perform tests in which both $T_e$ and $T_i$ are set to zero at the outer boundary. Even under that condition the boundary temperature remains above 100 eV, that is, above the range in which the rate coefficients vary, and no enhancement of the penetration was observed. We attribute this to the high edge temperature and the fast parallel transport in this discharge, which make it difficult to reduce the boundary temperature to the level at which the coefficients fall. In the edge energy balance of Table 4 the parallel conduction is the largest restoring contribution, and the net change of the two temperatures over the pulse remains at the level of tens of eV at the outer flux surfaces.